\documentclass[%
 aip,
 amsmath,amssymb,
 reprint,%
]{revtex4-1}

\usepackage{graphicx}% Include figure files
\usepackage{dcolumn}% Align table columns on decimal point
\usepackage{bm}% bold math
\usepackage{hyperref}

\usepackage[utf8]{inputenc}
\usepackage[T1]{fontenc}
\usepackage{mathptmx}
\usepackage{etoolbox}
\makeatletter
\def\@email#1#2{%
 \endgroup
 \patchcmd{\titleblock@produce}
  {\frontmatter@RRAPformat}
  {\frontmatter@RRAPformat{\produce@RRAP{*#1\href{mailto:#2}{#2}}}\frontmatter@RRAPformat}
  {}{}
}%
\makeatother
\begin{document}

\preprint{AIP/123-QED}

\title[WEEVIL]{WEEVIL:\\Design and Commissioning of an Ice-Sublimation Chamber and a Cryogenic Submillimeter Radiometer for Studying Rarefied Gas Plumes}
% Force line breaks with \\
\author{D. Belousov}
\email{dominik.belousov@unibe.ch}
\author{J. Fritschi}
\author{O. Mokhtari}
\author{L. L. Stöckli}
\author{D. Piazza}
\affiliation{
Space Research and Planetary Sciences, Physics Institute,
University of Bern, Sidlerstrasse 5, CH-3012 Bern, Switzerland
}

\author{A. Murk}
\affiliation{
Institute of Applied Physics, University of Bern,
Sidlerstrasse 5, CH-3012 Bern, Switzerland
}

\author{N. Thomas}
\affiliation{
Space Research and Planetary Sciences, Physics Institute,
University of Bern, Sidlerstrasse 5, CH-3012 Bern, Switzerland
}

\date{\today}% It is always \today, today,
             %  but any date may be explicitly specified

\begin{abstract}
We present WEEVIL (WatEr Emission of Vapor from Ice in the Laboratory), an experimental facility developed to study ice sublimation under laboratory conditions to support observations of comets and icy moons. The setup combines a submillimeter heterodyne radiometer and a high-vacuum cryogenic chamber, enabling spatially and temporally resolved monitoring of gas plumes formed during ice sublimation under controlled laboratory conditions. This paper describes the facility’s design and implementation. Furthermore, we present the commissioning of the experimental setup, including the first water-vapor spectra acquired during an ice sublimation experiment.

\end{abstract}

%\pacs{}% insert suggested PACS numbers in braces on next line

\maketitle %\maketitle must follow title, authors, abstract and \pacs 
% Body of paper goes here. Use proper sectioning commands. 
\section{Introduction}

Comets are considered among the most primitive bodies in the Solar System, preserving material from the early stages of planet formation \cite{Opik1973}. As they approach the Sun, increasing solar irradiation drives the sublimation of volatile ices, producing a gaseous coma that can be observed remotely from both ground- and space-based observatories. Observations of cometary volatiles provide valuable information on the composition and physical conditions of the solar nebula, and therefore offer important constraints on the processes that governed the formation and evolution of the Solar System. In addition, comets can be regarded as natural laboratories for investigating a broad range of fundamental physical processes, including rarefied gas dynamics, plasma interactions, thermodynamics, dust transport in gas flows, and heat and mass transfer in porous media \cite{Thomas2020}.

The Rosetta mission provided an unprecedented characterization of the morphology, composition, and coma dynamics of comet 67P/Churyumov–Gerasimenko (hereafter 67P) \citep{Glassmeier2007, Thomas2020}. Several instruments investigated the innermost coma, including the Microwave Instrument for the Rosetta Orbiter (MIRO), which used millimeter- and submillimeter (sub-mm) radiometers to measure rotational transitions of volatile species \cite{Gulkis2007}. The initial analysis of the MIRO data used a parametric one-dimensional coma model \cite{Biver2015}, showing significant spatial variability in the gas emitted from the nucleus. Subsequent studies introduced three-dimensional Direct Simulation Monte Carlo (DSMC) models that incorporated the complex nucleus shape, heterogeneous surface activity, and more realistic boundary conditions for gas emission \cite{Marschall2019}. These studies showed that both nucleus morphology and the physical properties of the surface and subsurface layers strongly influence the observed coma structure. One of MIRO's principal scientific objectives was to investigate the processes controlling inner-coma development \cite{Gulkis2007}. The instrument provided an extensive data set of gas profiles at altitudes from 100 m to 10 km above the nucleus surface \cite{Marschall2019}. At the same time, high-resolution imaging of the dust coma obtained by the Optical, Spectroscopic, and Infrared Remote Imaging System (OSIRIS) inferred a highly complex gas-flow field \cite{Keller2007}. The near-absence of highly reflective ices at 67P and the high surface temperatures seen at other comets (e.g., 1P/Halley; see \citet{Emerich1987}) also suggest that volatile sublimation occurs primarily close to, but just below, the visible surface. As a result, heterogeneous surface activity, possible gas transport through a porous surface layer, and subsequent rarefied gas dynamics make interpreting Rosetta observational data highly challenging and ambiguous, with the degeneracy of equally well-fitting model parameters evident \citep{Marschall2020}.

Numerical models describing gas transport through porous media have advanced considerably, including applications to cometary nuclei \cite{Dullien1991, Skorov2011}. More recently, models have been developed that couple gas diffusion through porous dust layers with its subsequent expansion into the inner coma, providing a more self-consistent description of cometary activity \citep{Thomas2020, Christou2020, Pinzon_Rodriguez2021, Mokhtari2024}. As a result, current numerical models incorporate increasingly sophisticated physics, including subsurface gas transport, heterogeneous material properties, and complex gas–dust interactions. However, these models include free parameters and certain simplifications. In particular, boundary conditions at the sublimation interface often rely on the Hertz–Knudsen equation for the sublimation flux. At the same time, emitted molecules are assumed to follow a half-Maxwellian velocity distribution \cite{Huebner2006}. The exact physics governing gas-molecule interactions with the dust-grain structure within the surface dust layer remains poorly constrained, leaving free parameters in gas-dynamic models, such as the sticking and accommodation coefficients.

Despite these advances in numerical models, experimental validation has not kept pace. Existing laboratory studies have focused mainly on integrated quantities such as sublimation rates, gas production, chamber pressure, and dust ejection, while spatially resolved direct measurements of expanding gas flow remain scarce \citep{Lmmerzahl1995, Gundlach2011, Bischoff2019, Kossacki2022, Kossacki_2026}. Radiometry is a promising technique to fill this gap. Radiometers are widely used for remote sensing of planetary atmospheres, icy moons, and small bodies \citep{Gulkis2007, deGraauwTh2010, Nordh2003, Hartogh2026}. By observing molecular transitions at microwave and sub-mm wavelengths, radiometers can retrieve gas temperature, density, velocity, and molecular composition, including isotopic abundances and their spatial distributions. Thus, applying this technique under controlled laboratory conditions allows gas profiles to be obtained above sublimating ice. By reproducing comet-relevant conditions in a controlled environment, such experiments enable the validation of radiative-transfer and gas-dynamics models, as well as the assumptions adopted in the retrieval of physical gas properties. They also support the interpretation of data from spaceborne radiometers (e.g., MIRO, Submillimeter Wave Instrument (SWI)\cite{Hartogh2026}), help quantify the capabilities and limitations of sub-mm radiometry for cometary observations, and contribute to the development of future microwave instruments.

Here, we present a laboratory setup called WEEVIL (WatEr Emission of Vapor from Ice in the Laboratory) designed to study water-ice sublimation and gas properties above the sublimating surface using sub-mm heterodyne radiometry. This paper describes the scientific objectives and setup requirements, followed by a detailed description of the setup itself, including the sub-mm heterodyne radiometer. We discuss commissioning and first spectral observations before briefly describing our plans for forthcoming measurements. 

\section{Scientific objectives}

The primary scientific objectives of the WEEVIL setup are:

\begin{enumerate}
    \item \textbf{Improve gas-dynamics modeling}
    \begin{itemize}
        \item Measure the spatial distribution of gas temperature, density, and velocity above sublimating icy surfaces.
        \item Determine appropriate gas boundary conditions at the sublimating surface from laboratory observations.
        \item Compare laboratory observations with DSMC simulations and numerical models of transport in porous media.
    \end{itemize}
    \item \textbf{Validate microwave retrieval techniques}
    \begin{itemize}
        \item Investigate how accurately sub-mm spectra recover gas physical properties under comet-relevant conditions. 
        \item Quantify uncertainties associated with radiative transfer retrieval algorithms.
    \end{itemize}
    \item \textbf{Support past and future missions to comets and icy moons}
    \begin{itemize}
        \item Assess the capabilities and limitations of sub-mm radiometers for observations of cometary activity. 
        \item Use laboratory measurements to improve the interpretation of Rosetta/MIRO  and JUICE/SWI observations.
        \item Provide experimental data for instrument design, observation strategies, and calibration of future missions to comets.
    \end{itemize}
\end{enumerate}

\section{\label{sec:setup_requirements} Setup requirements}

The WEEVIL facility is designed to investigate the physical properties of gas released during water-ice sublimation under comet-relevant conditions. The primary observables are the profiles of gas density and temperatures (rotational and translational) as functions of height above the sublimating surface.

Measurements are performed in a limb-viewing geometry in which the radiometer beam crosses the expanding gas plume parallel to the sample surface. This geometry is particularly advantageous for molecular spectroscopy because the long optical path through the gas plume enhances line contrast against the continuum background, while measurements at different heights probe the plume's vertical structure. Spaceborne sub-mm radiometers, including MIRO and SWI, use similar limb-viewing observations to probe gas properties at different altitudes above 67P and Jupiter's moons, respectively \citep{Rezac2019, Hartogh2026}. A vertical scanning mechanism is therefore required to measure gas properties at different heights above the surface with high spatial resolution, resolving gradients associated with gas acceleration and collisional processes within the Knudsen layer.

The radiometer must provide high spectral resolution to accurately determine the shapes of narrow emission lines and achieve a high signal-to-noise ratio (SNR) with short integration times (minutes) to investigate transient phenomena and the time-dependent evolution of the gas flow.

The experimental environment should minimize systematic effects unrelated to ice sublimation. In particular, the measured signal should predominantly originate from gas molecules released from the sample surface, while the contribution of water molecules desorbed from the chamber walls should be minimized to avoid bias in the retrieved gas properties. The chamber background pressure should be below $10^{-6}$ mbar prior to sublimation experiments. 

The ice sample temperature should be actively controlled in the range of 170 to 220 K (to initiate sublimation of water ice), with long-term stability better than $\pm 0.5$ K over several hours, depending on the sublimation rate. In addition, the setup should enable controlled irradiation of the sample from above using a solar simulator to simulate solar heating.  

Gas-dynamics models should provide a priori information on gas density, temperature, and velocity profiles for use in radiometric data retrievals \cite{Rodgers2000}. Gas observations with additional sensors, such as pressure gauges or quartz crystal microbalances (QCMs), should provide complementary constraints on the retrieved gas properties.  

\section{Design}

\subsection{\label{sec:vacuum_chamber} Vacuum chamber}

\begin{figure}[t]
    \includegraphics[width=0.4\textwidth]{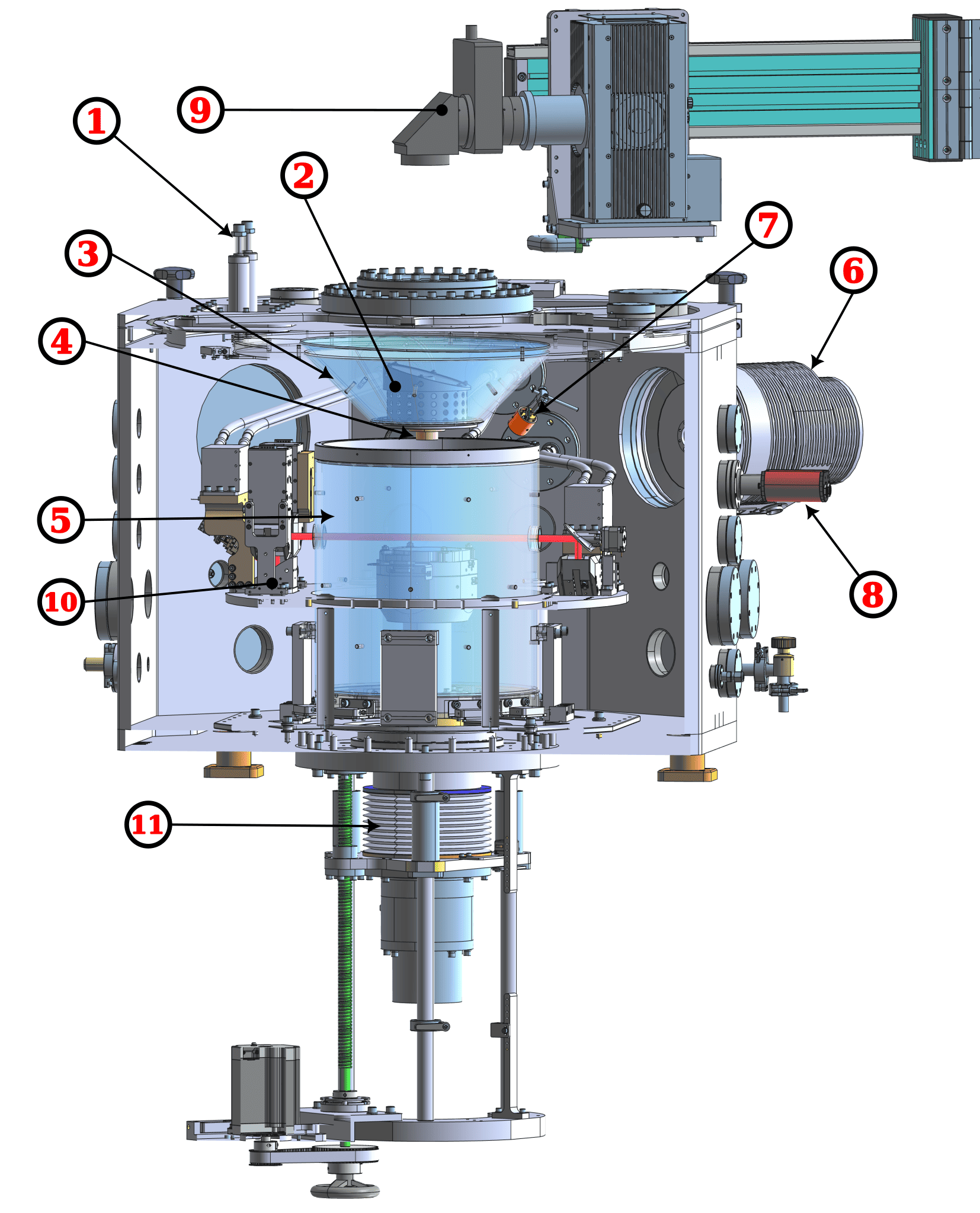}
    \caption{\label{fig:chamber_drawing} Technical drawings of WEEVIL setup. 
    1: LN$_{2}$ feedthroughs;
    2: angled window attached to cooled collar;
    3: top shroud;
    4: quartz crystal microbalance sensor;
    5: base shroud (Sec.~\ref{sec:cooling_subsystem});
    6: turbomolecular pump;
    7: IR thermopile-based temperature sensor;
    8: pressure gauge;
    9: solar simulator;
    10: radiometer (Sec.~\ref{sec:radiometer_subsystem});
    11: cryocooler with the scanning mechanism (Sec.~\ref{sec:scanning_subsystem}).}
\end{figure}

\begin{figure*}[t]
    \centering
    \begin{minipage}{0.49\textwidth}
        \raggedright
        (a)\par
        \includegraphics[width=\linewidth]{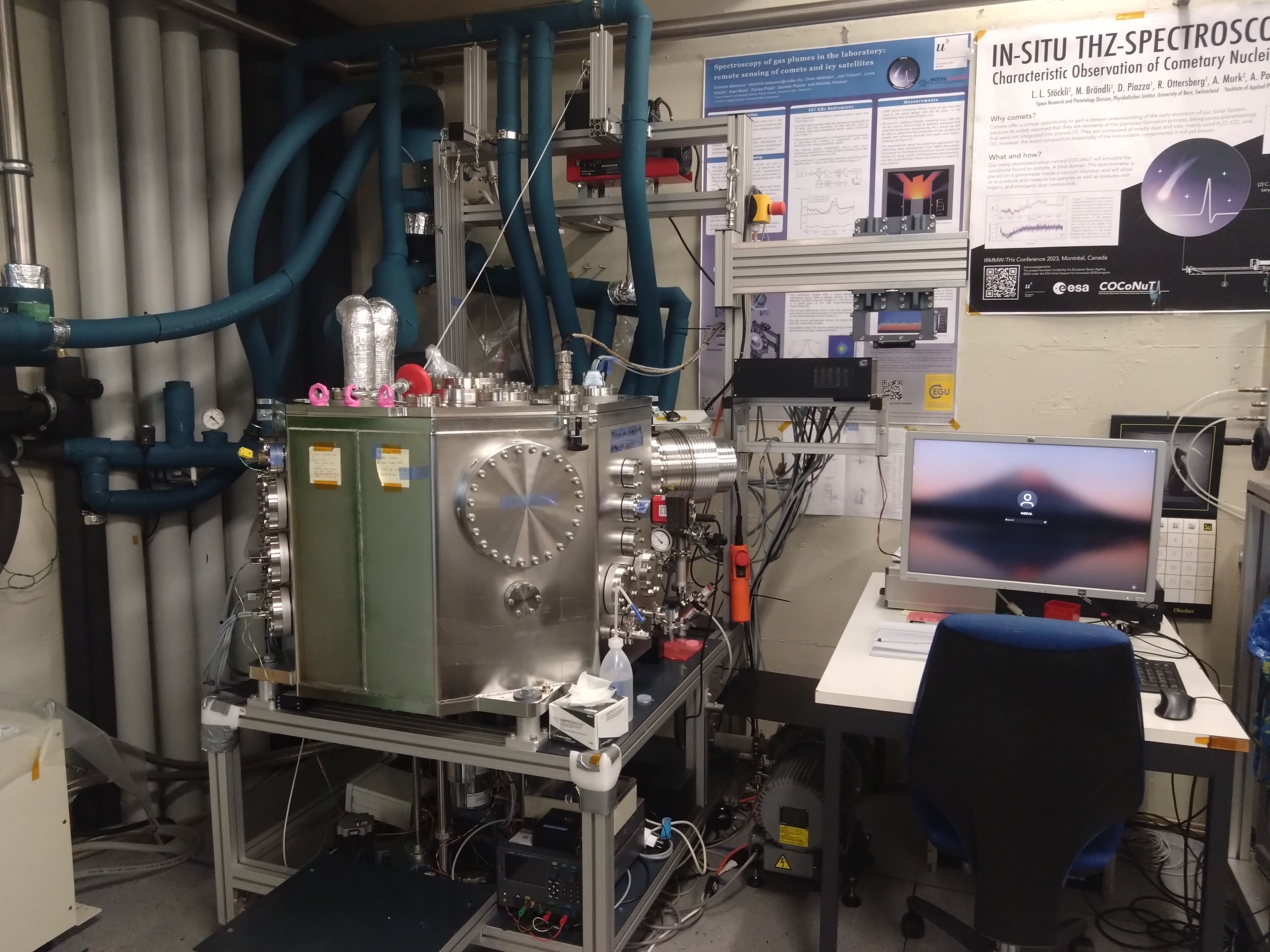}
    \end{minipage}\hfill
    \begin{minipage}{0.49\textwidth}
        \raggedright
        (b)\par
        \includegraphics[width=\linewidth]{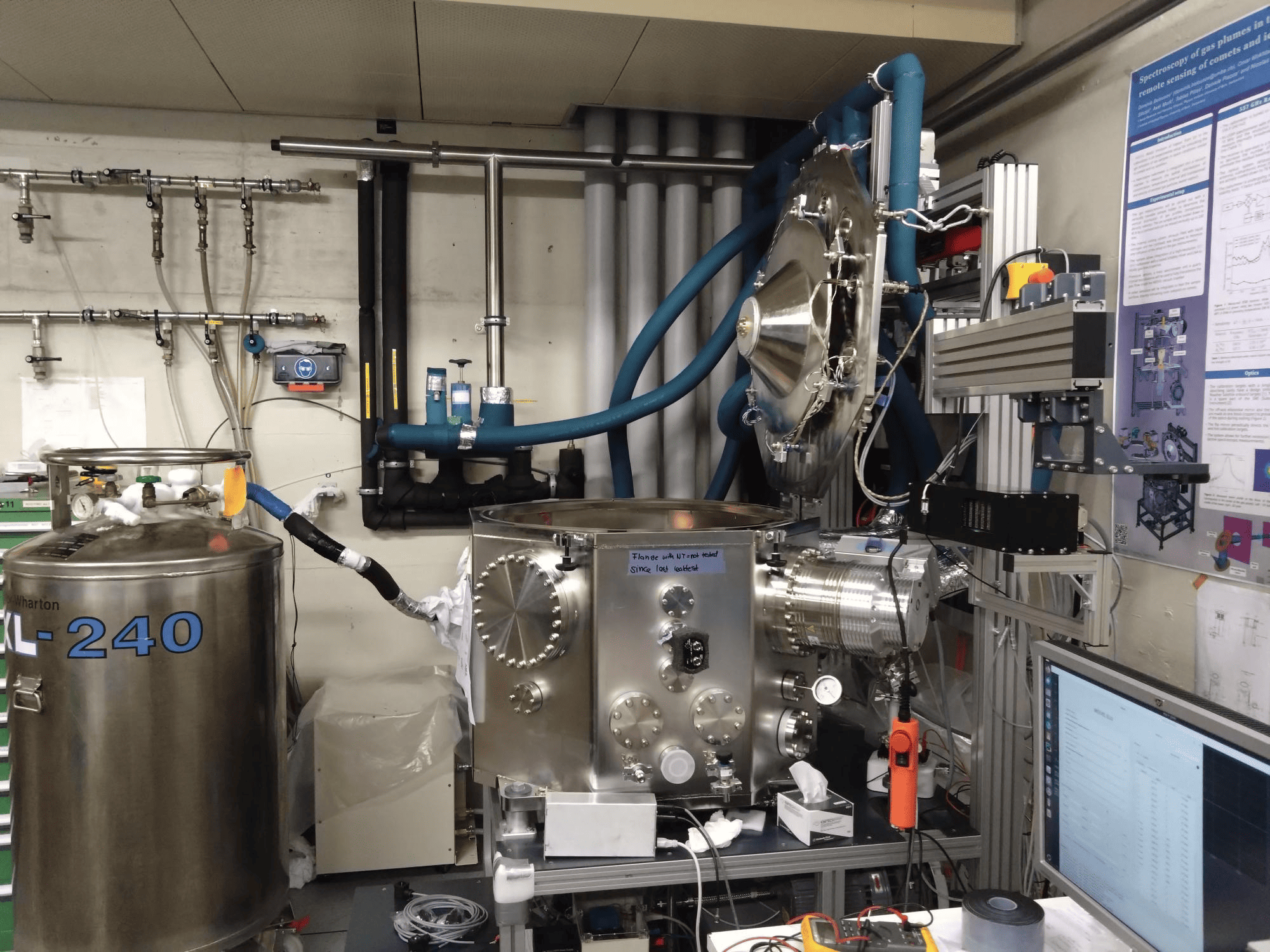}
    \end{minipage}

    \caption{\label{fig:chamber_in_lab}
    The WEEVIL setup installed in the laboratory.
    (a) Front view with the vacuum-chamber lid closed.
    (b) Side view with the lid open, showing the lid-opening mechanism
    and the microbalance sensor mounted on the bottom ring of the top
    shroud. An external LN$_2$ reservoir is connected to the setup's
    LN$_2$ inlet.}
\end{figure*}

The main part of the WEEVIL setup is a vacuum chamber with an integrated active cooling system. The chamber serves two primary purposes: it reproduces comet conditions and provides active cooling for the sub-mm radiometer (details are in Sec.~\ref{sec:radiometer_subsystem}). Fig.~\ref{fig:chamber_drawing} shows a technical drawing of the complete experimental setup, and Fig.~\ref{fig:chamber_in_lab} shows a photograph of the assembled system. The octagonal chamber geometry provides flat external surfaces that simplify flange placement. The chamber footprint is roughly 900 mm $\times$ 900 mm with a height of 600~mm, yielding an internal volume of $\approx$ 400~liters (0.4~m$^3$) and a total weight of $\approx$ 900~kg. The initial rough pumping is performed using a Leybold Scrollvac SC 30D oil-free scroll pump with a nominal pumping speed of 30~L/s, reducing the chamber pressure to $<1{\times}10^{-2}$ mbar. After initial pumping, the Turbovac MAG W 700 iP turbomolecular pump (TMP) with a pumping speed of 590~L/s can be turned on to achieve a high vacuum ($10^{-3}-10^{-7}$~mbar). The chamber pressure is monitored using a Pfeiffer HPT 200 AR pressure gauge with an accuracy of 10$\%$ from $10^{-8} $ to $10^{1}$~mbar, while the foreline pressure between the roughing and turbomolecular pumps is measured with a Pfeiffer PTR 280 pressure gauge.  

High-precision direct measurement techniques, such as a quartz crystal microbalance (QCM), are used to characterize the gas flux/production rate from the ice sample \cite{Galli2017, Grundy2024}. The QCM directly measures the mass flux inside the vacuum chamber, complementing the radiometer remote-sensing observations. A TQCM Mark~10 microbalance sensor (QCM Research, USA) is mounted on an aluminum plate attached to the bottom of the top shroud (Fig.~\ref{fig:chamber_drawing}) and faces the ice sample. This configuration ensures stable thermal conditions for the QCM sensor while providing an optimal viewing geometry to probe gas plume fluxes.

\subsubsection{\label{sec:cooling_subsystem} Cooling subsystem}
The system is designed to minimize gas flow reflected from the chamber walls at room temperature, thereby reducing interference with gas originating from the sublimating ice sample. Cold shrouds inside the vacuum chamber limit back reflection of gas particles from the warm (room temperature) chamber walls. The shrouds are made of electropolished stainless steel, which facilitates regeneration by desorbing water accumulated on their surfaces during sublimation experiments, and also provides smooth, well-defined surfaces for the gas-flow boundary conditions. The shroud and chamber sizes were determined by combining DSMC gas simulations of preliminary designs with manufacturability constraints. The main limitation to achieving ideal undisturbed gas flow arises from the window required for the solar insolation simulator. The design includes a tilted window to partially deflect the gas particles reflected from the window to the cold shroud (Fig.~\ref{fig:chamber_drawing}). Fig.~\ref {fig:dsmc_model} shows the dimensions of the gas flow chamber. In addition, plates of different materials and shapes can be attached to the top shroud. The reference experiment will use an aluminum cold plate. Because the solar simulator beam is blocked by a plate, heaters inside the sample holder heat the sample. 

\begin{figure}
\includegraphics[scale=0.4]{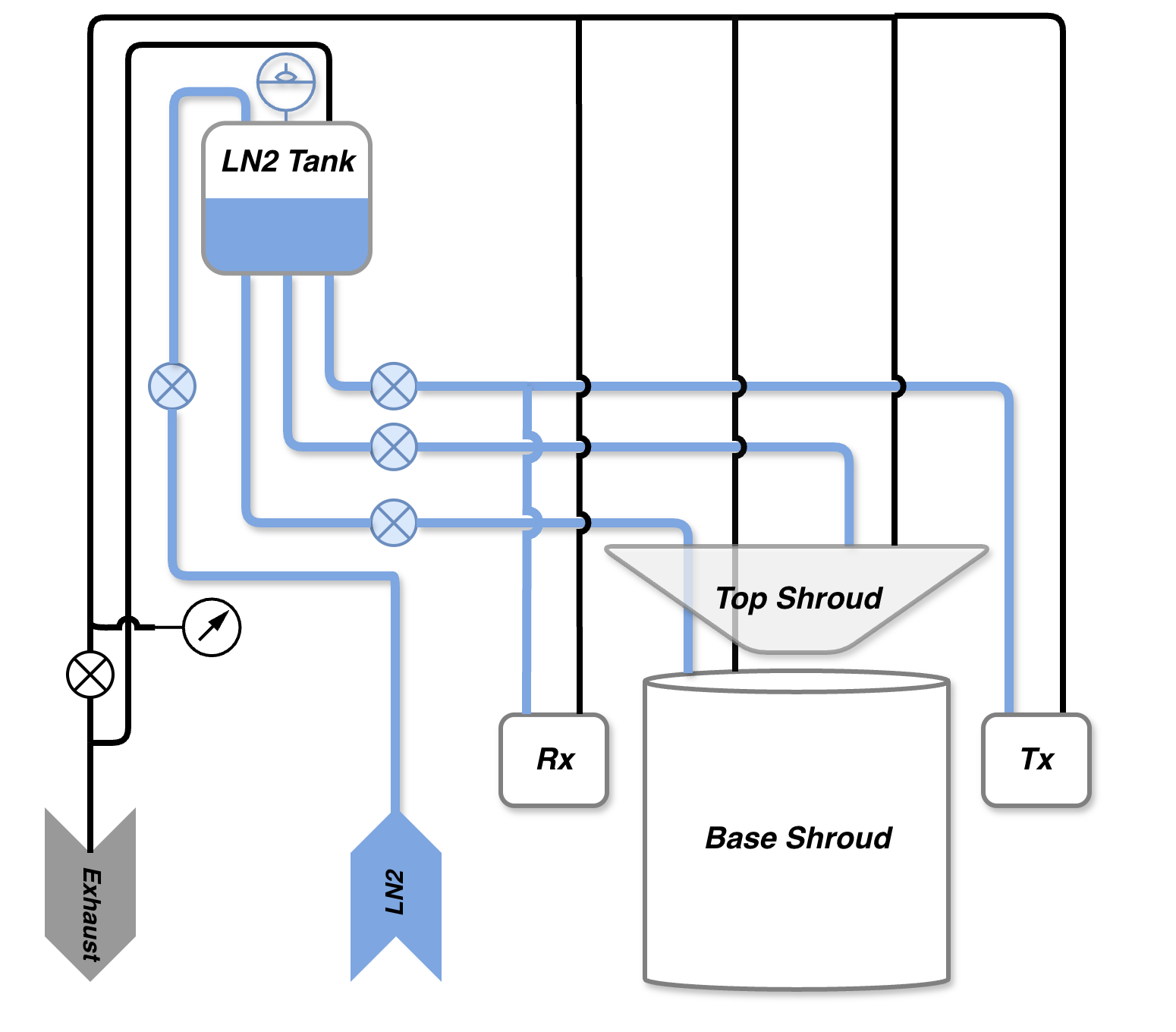}
\caption{\label{fig:ln2_system} Block diagram of the liquid nitrogen cooling system.}
\end{figure}

The integrated cooling system consists of stainless steel shrouds that are filled with liquid nitrogen (LN$_2$) during operation. The system comprises four separate LN$_2$ reservoirs: the top shroud, the base shroud (both act as a cryogenic trap for gas condensation and have a volume of 2~L), and two radiometer cooling blocks (both 0.2~L). The stainless steel shrouds (top and base) were electropolished to minimize surface roughness, which is critical for achieving zero-slip boundary conditions during gas-flow simulations on the cold surfaces. Each cooling circuit is replenished independently using automatically controlled valves, enabling stable long-term operation without interrupting the experiment. The LN$_2$ reservoirs are supplied through a 7~L main LN$_2$ tank equipped with an LN$_2$-level sensor (Consarctic NLM-1). This buffer tank stabilizes the LN$_2$ level and the temperature of the individual circuits, helping limit the temperature fluctuations that affect radiometer operation. To further improve the radiometer's temperature stability, a 10~W resistive heater is integrated into the ellipsoidal mirror block (Sec.~\ref{sec:radiometer_subsystem}) and controlled by a proportional–integral–derivative (PID) controller. Electrical high-duty cartridge heaters are attached to the cooling shrouds to accelerate warm-up after the experiment and to facilitate the removal of condensed water and other volatiles from the shroud surfaces. The LN$_2$ distribution system also incorporates a pressurizing valve that restricts the gaseous nitrogen exhaust line, increases the pressure in the LN$_2$ supply line, and enforces the filling of the cooling circuit. Fig.~\ref {fig:ln2_system} shows a block diagram of the LN$_2$ system. 

The ice sample must be maintained at a stable and well-controlled temperature throughout an experiment. To reproduce a wide range of comet-relevant thermal conditions, the sample holder is actively temperature-controlled using a closed-cycle helium cryocooler consisting of a Sumitomo F-70 compressor and a CH-110 cold head. Because the cryocooler's cooling power cannot be directly regulated, high-duty heater cartridges establish and maintain the desired sample temperature. Two 160~W heaters mounted in the sample holder bore holes are controlled by a PID controller, which compensates for the continuous cooling provided by the cryocooler and provides stable temperature regulation ($\pm$0.5~K) in the range of 40--220~K throughout the experiment, using an approach similar to that described by \citet{Stckli2025}.

The sample holder is circular, with a diameter of 106~mm and a thickness of 10~mm. It is fully interchangeable with the sample holder used in the SCITEAS-2 setup \citep{Ottersberg2025}, which has also been used for cryogenic investigations of cometary simulants.

The cryocooler cold head temperature is measured using a Lake Shore DT-670D silicon diode. The temperature of the various components of the experimental setup is monitored using Pt100 and Pt1000 resistance temperature detectors (RTDs), or BAS16 silicon diodes mounted in custom-made temperature-sensor holders \cite{Stockli2024}. The surface temperature of the ice sample is measured using an infrared thermopile-based temperature sensor \cite{Ottersberg2025}, and its bulk temperature can be measured using a temperature sensor frozen into the ice. Alternatively, the sample holder temperature can be measured and used as an approximation of the ice temperature. 

\subsubsection{\label{sec:scanning_subsystem} Scanning subsystem}
Gas measurements use a vertically movable sample holder to determine the vertical distribution of gas profiles (temperature, density, velocity) in the limb observation geometry. The cryocooler with the ice sample on top can be lowered vertically to 200~mm. The system's "zero position" is defined at the exact point where the sub-mm measurement beam passes directly over the surface of the ice sample. From this baseline, the sample can be lowered, systematically increasing the distance between the target material and the fixed measurement axis to capture vertical plume profiles.

\begin{figure}
\centering
\includegraphics[scale=0.05]{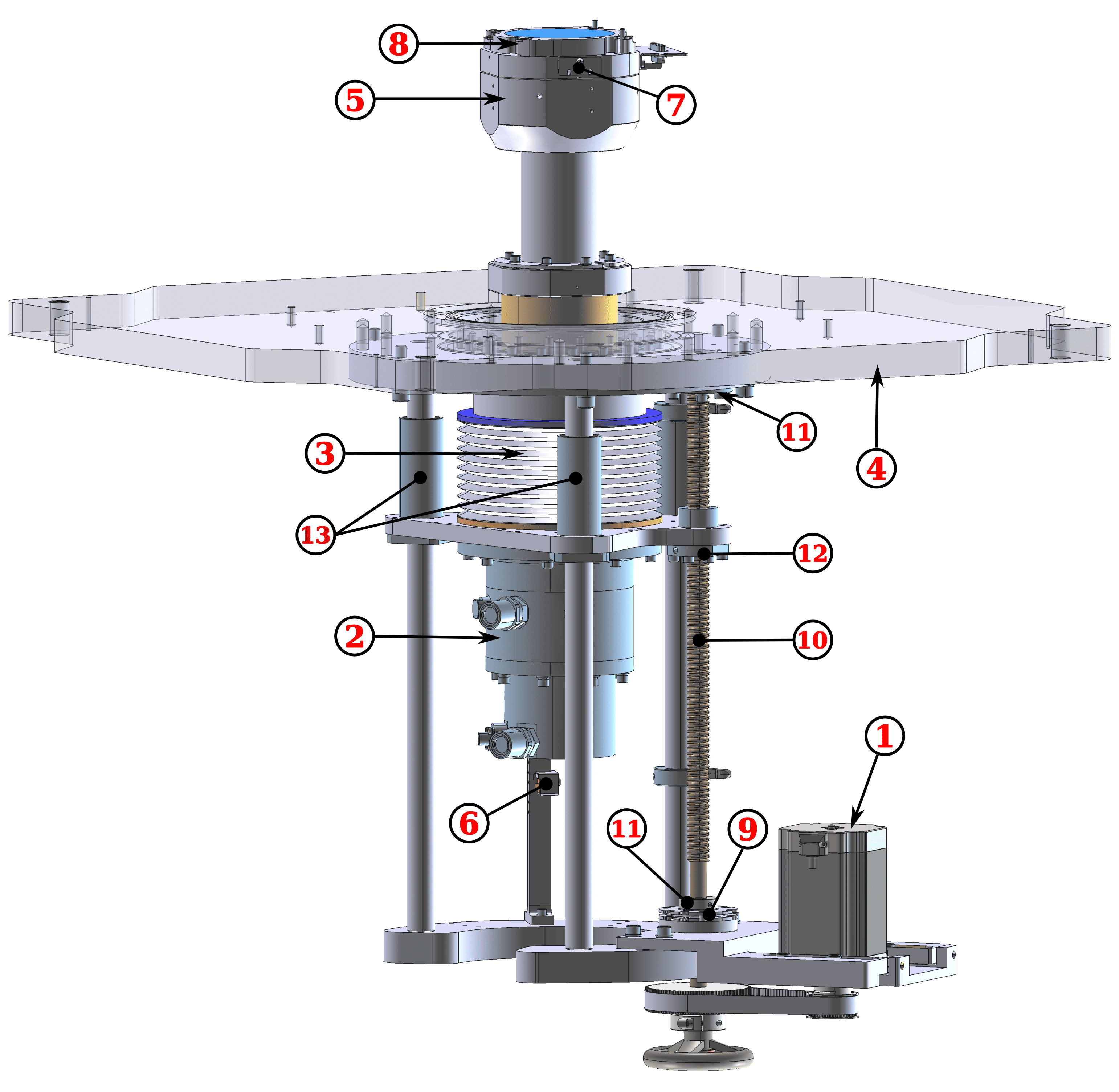}
\caption{\label{fig:cryocooler_stage}
Scanning linear-stage mechanism with the installed cryocooler.
1: stepper motor;
2: cryocooler;
3: vacuum bellows;
4: vacuum-chamber bottom plate;
5: cryohead adapter;
6: Hall-effect position sensor;
7: high-duty cartridge heaters;
8: sample holder;
9: spring brake;
10: spindle;
11: ball bearings;
12: ball screw nut;
13: linear bearings.}
\end{figure}

A flexible vacuum bellow connects the translation system to the chamber vacuum flange, allowing the sample holder to move vertically while maintaining a vacuum seal (Fig.~\ref{fig:cryocooler_stage}). Keeping the radiometer fixed preserves the receiver's optical alignment while allowing the gas plume to be sampled at different altitudes. The moving cryohead requires a stepper motor, ball screws, a spindle, brakes, and positioning sensors. The cold section of the cryohead remains in vacuum, while the motor and lubricated ball screws are outside the vacuum chamber. This prevents motor heat from disrupting internal thermal zones and eliminates lubricant-related vacuum contamination. A Sanyo Denki SM2863 stepper motor drives the vertical scanning mechanism, with TMC5160 motor drivers integrated into the Electronics Unit (ELU). An electromagnetic brake locks the spindle in position when the motor is unpowered, preventing drift under vacuum or gravitational loads. The external stage utilizes three linear guide rods paired with ball screws to ensure highly stable, rigid vertical translation. As a result, the scanning mechanism ensures precise, reproducible sample positioning.

\subsubsection{\label{sec:electronics_software} Electronics and Software subsystem}

Operation of the vacuum chamber - including control of pumps, heaters, LN$_2$ valves and motors, as well as the acquisition of housekeeping (HK) data - is managed by the custom-made Electronic Local Unit (ELU). \citet {Stckli2025} provides a detailed description of the ELU. Briefly, the ELU is built around an STM32F405 ARM microcontroller and features a modular architecture with multiple extension boards for controlling various subsystems of the experimental setup. The control PC transmits commands to the ELU via an RS-232 communication protocol, including HK data commands. HK data are continuously transmitted to a central server and visualized in Grafana, a web-based platform for real-time monitoring of experimental parameters.

The instrument control software was developed in Python and provides unified control of both the radiometer and the ELU, including data recording, HK monitoring, and command sending. Spectra are recorded in HDF5 format. The software follows a modular actor-based architecture, in which each hardware subsystem is implemented as an independent actor that communicates exclusively through a message-passing interface \cite{Agha1999}. This design simplifies management of multiple instruments, facilitates integration of additional hardware modules, and improves overall control-system stability. The software supports both interactive manual operation (using a GUI) and fully automated scripts. This approach gives the operator additional flexibility during instrument development, commissioning, and routine operation.  
 
\subsection{\label{sec:radiometer_subsystem} Radiometer}

\begin{figure*}[t]
    \centering
    \begin{minipage}{0.45\textwidth}
        \raggedright
        (a)\par
        \includegraphics[width=\linewidth]{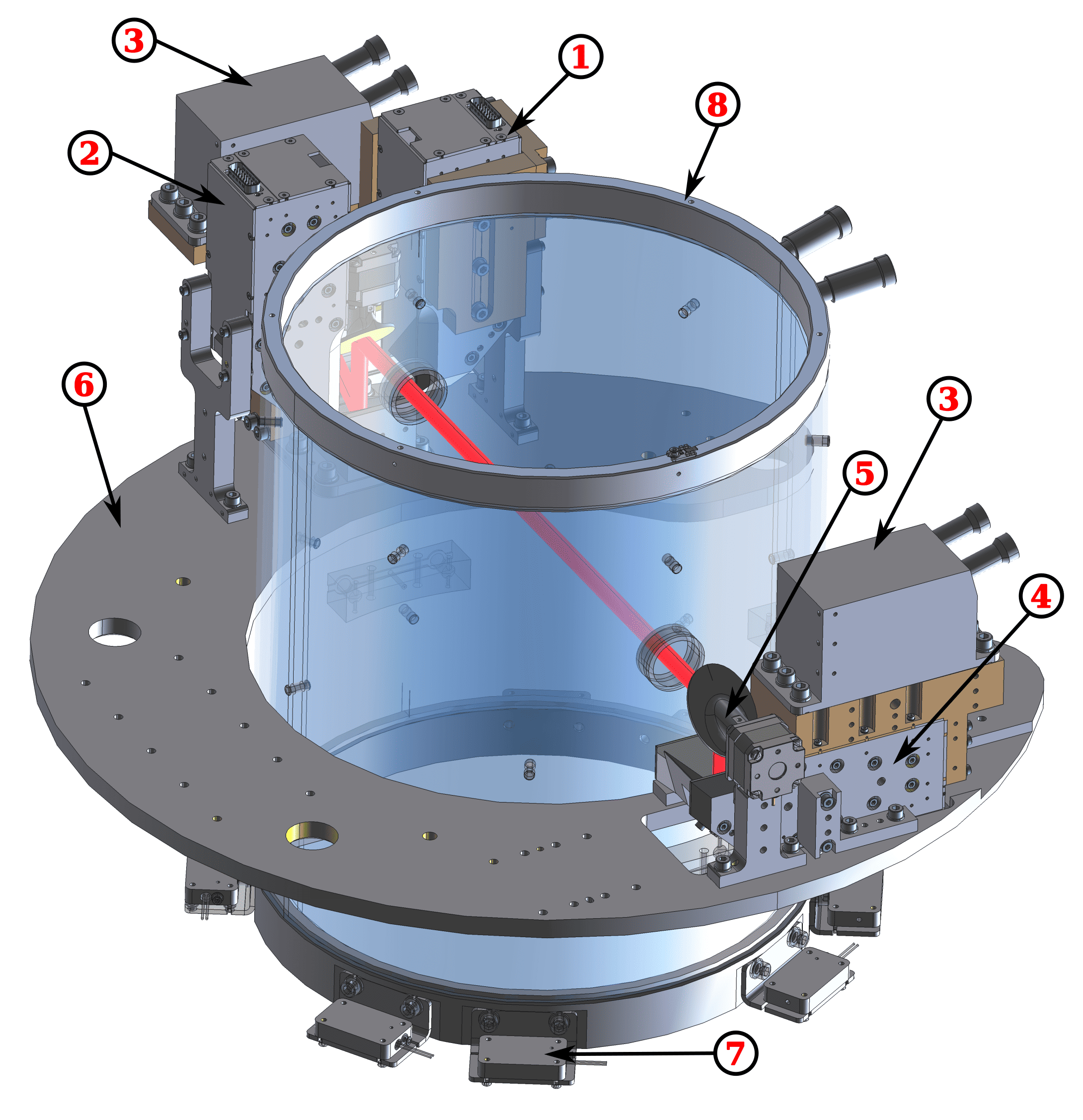}
    \end{minipage}\hfill
    \begin{minipage}{0.45\textwidth}
        \raggedright
        (b)\par
        \includegraphics[width=\linewidth]{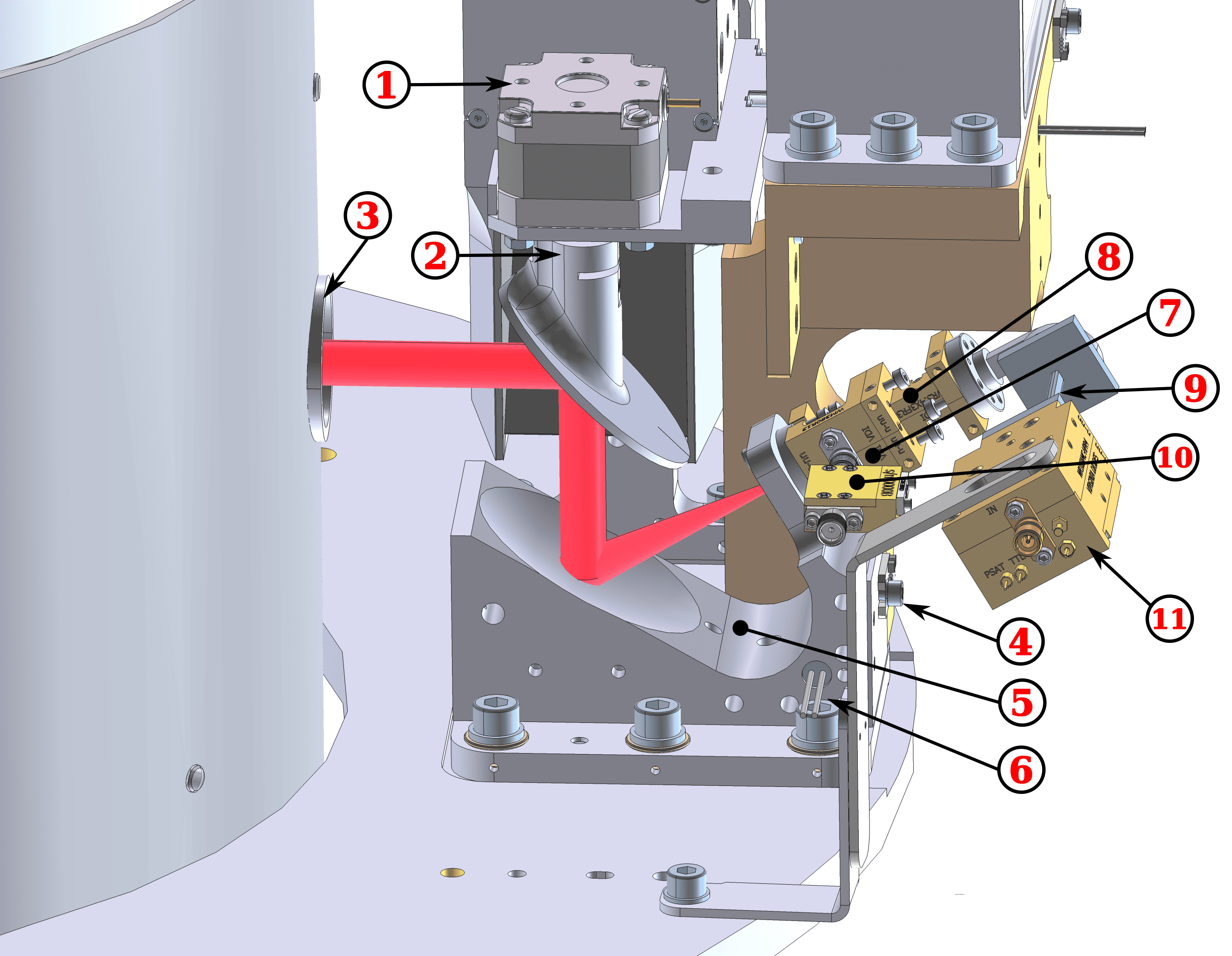}
    \end{minipage}

    \caption{\label{fig:radiometer_layout}
    The 557~GHz radiometer of the WEEVIL setup. The radiometer beam is
    shown in red.
    (a) Radiometer bench surrounding the base shroud:
    1: cold target; 
    2: hot target; 
    3: LN$_2$ volumes;
    4: background target; 5: background flip mirror;
    6: radiometer table; 
    7: base-shroud heaters; 8: base shroud.
    (b) Close-up view of the radiometer receiver:
    1: stepper motor; 
    2: calibration flip mirror; 
    3: shroud aperture;
    4: PT1000 temperature sensor; 
    5: mirror block comprising the
    ellipsoidal mirror and antenna holder; 
    6: radiometer heater;
    7: subharmonic mixer; 8: frequency tripler;
    9: stainless-steel WR10 waveguide; 
    10: low-noise amplifier (LNA);
    11: Active Multiplier Chain (AMC).}
\end{figure*}

A Dicke-switch heterodyne radiometer operating primarily near 556.936~GHz (ortho-water ground-state rotational transition $1_{01} \to 1_{10}$) has been developed to measure water vapor column densities, production rates, gas temperatures, and outflow velocities. The radiometer concept draws on both the laboratory experimental setup described in \citet{Auriacombe2022} and the design of spaceborne sub-mm radiometers \citep{Gulkis2007, Hartogh2026}. A key improvement over \citet{Auriacombe2022} is integrating the radiometer inside the vacuum chamber. This configuration eliminates atmospheric absorption along the receiver optical path, enables observations of optically thick spectroscopic lines, and reduces receiver noise through active cryogenic cooling of radiometer components. Furthermore, removing the vacuum windows from the receiver beam eliminates window-induced standing waves and associated spectral baseline ripples, and minimizes optical losses. The radiometer system comprises a heterodyne receiver for detecting submillimeter-wave radiation together with a set of optical components to form the scanning near-field beam passing through a gas plume and enabling receiver calibration using hot and cold blackbody targets (Fig.~\ref{fig:radiometer_layout}). In addition, the radiometer layout allows the integration of a transmitter for active measurements of weak molecular transitions, such as the H$_2^{18}$O line at 547.7~GHz.

\subsubsection{\label{sec:level3}Heterodyne DSB receiver}

\begin{figure*}
\includegraphics[scale=0.50]{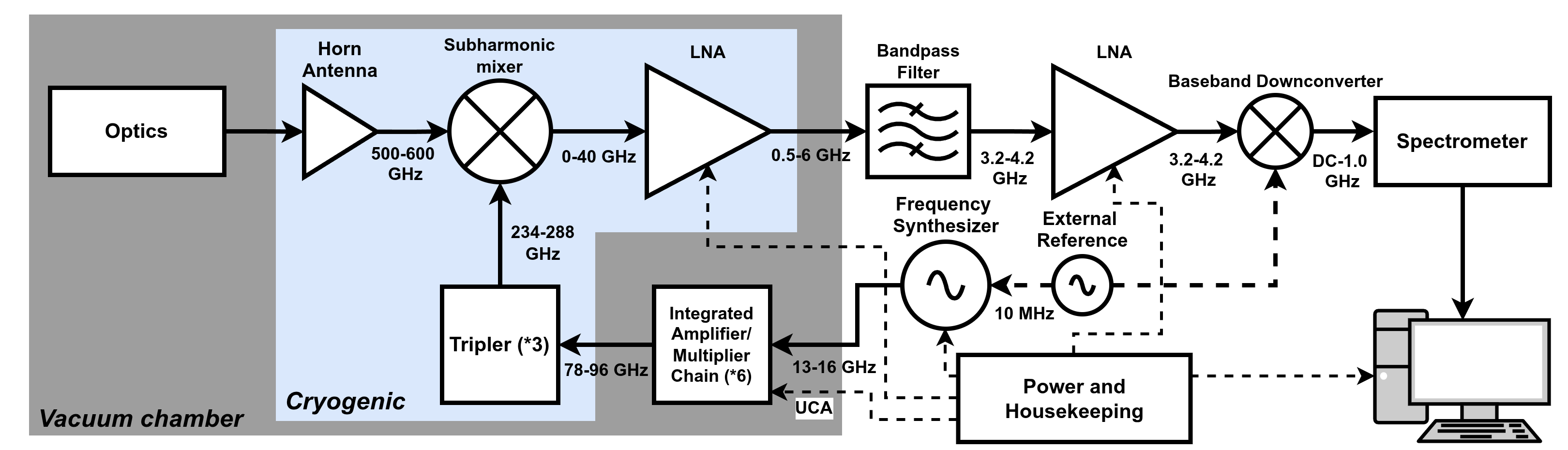}
\caption{\label{fig:radiometer_block_diagram} Block diagram of the heterodyne receiver.}
\end{figure*}

The radiometer employs a heterodyne architecture based on a GaAs Schottky diode subharmonic mixer, which combines the radio-frequency (RF) signal with the second harmonic of the local oscillator (LO) signal and converts it to an intermediate-frequency (IF) band. The mixer is sensitive to RF signals of the upper and lower sidebands relative to the LO frequency, resulting in double-sideband (DSB) detection, similar to the MIRO instrument \citep{Gulkis2007}.

Fig.~\ref{fig:radiometer_block_diagram} shows the block-diagram of the sub-mm receiver. The LO chain driving the subharmonic mixer (WR1.9SHM, Virginia Diodes Inc., VDI, USA) consists of an active multiplier chain (WR10AMC-I, VDI), a broadband tripler (WR3.4X3FUHP, VDI), and a microwave frequency synthesizer (SMA100B Signal Generator, Rohde$\&$Schwarz). The IF signal is first amplified by a low-noise amplifier (LNA) (LNF-LNC1.5, Low Noise Factory) that covers the 1.5–6~GHz band with a gain ($G$) of 35~dB. It then passes through a 3.7 GHz band-pass filter (bandwidth ($B$) = 1.0 GHz) and is further amplified by a Miteq LNA ($B$ = 2-5 GHz, $G$ = 57 dB). A USRP N210 software-defined radio ($B$ = 25 MHz) digitizes the amplified IF signal and transmits it to a PC, where it is processed in real time with a Fast Fourier Transform (FFT) using the GNU Radio software framework \cite{Holland2015}. 

The subharmonic mixer, the LO tripler, and the IF LNA are connected to a copper block that is actively cooled with the LN$_2$ block to approximately 100~K. In contrast, the active multiplier chain (AMC) must operate close to room temperature and is therefore thermally isolated from the cooled tripler by a 50~mm-long stainless-steel WR10 waveguide. The WR1.9SHM mixer requires an LO drive power of approximately 1--5~dBm for optimal operation. Generating this power at submillimeter wavelengths requires the AMC, which contributes significantly to the radiometer power consumption and heat dissipation. During operation, the AMC consumes approximately 4.5--5 W of electrical power, most of which is ultimately dissipated as heat. To maintain a safe operating temperature under vacuum conditions, this heat is carried away through a copper thermal strap connected to the LN$_2$-cooled radiometer block. During cooling, a PID-controlled resistive heater maintains the AMC near room temperature.

To reduce the conductive heat load transmitted through the IF coaxial cable connecting the vacuum feedthrough to the cryogenic LNA, the coaxial adapter is thermally anchored to a copper block. This thermal intercept diverts heat conducted from the vacuum chamber wall to the copper block, thereby reducing the heat load on the LNF LNA. The LO chain provides frequencies between 220 and 330 GHz for pumping the subharmonic mixer. The LO power is adjusted using the AMC user controller attenuation (UCA), allowing tuning to achieve the radiometer receiver's highest sensitivity at frequencies from 500 to 600 GHz. A Keysight E36312A programmable DC power supply provides low-noise bias voltages for the radiometer components.

\subsubsection{\label{sec:optics} Optics}

Fig.~\ref{fig:radiometer_layout} shows the radiometer optical layout. The optical bench uses a single, removable plate that enables precise optical alignment at room temperature and maintains alignment after cryogenic shrinkage. A corrugated feedhorn couples the incoming RF radiation to a mixer waveguide. The receiver is sensitive to a single linearly polarized component of the incoming radiation. The 600 GHz corrugated feedhorn is a spare part developed for the SWI instrument \cite{Kotiranta2018} and was manufactured by Thomas Keating Ltd. (UK). The feedhorn beam is further directed to an off-axis ellipsoidal mirror with an incident angle of 30$^\circ$. This geometry minimizes the cross-polarization and beam aberrations introduced by the off-axis mirror. The feedhorn holder and ellipsoidal mirror are machined as a single gold-plated copper assembly to minimize relative misalignment caused by thermal contraction during cooling. The ellipsoidal mirror focuses the beam onto the center of the base shroud, producing a beam diameter of approximately 10 mm at the intensity level $-10$ dB (Fig.~\ref{fig:beam_profile_grasp}). A flip mirror, driven by a stepper motor, alternately directs the radiometer view toward a base shroud with 30 mm circular apertures or to a calibration target. During observations, the radiometer periodically performs a calibration cycle by observing the cold and hot calibration targets. These reference measurements compensate for long-term variations in receiver gain and system noise, ensuring stable and calibrated spectroscopic measurements. The dimensions of the mirrors, targets, and apertures in the base shroud used for gas observations were optimized using GRASP 10.5.0 (TICRA, Denmark), commercial software for reflector-antenna and quasi-optical system simulations, to minimize sidelobe levels and maximize beam efficiency. In addition, the cooled shrouds surrounding the optical path provide a low-temperature radiative background, reducing the contribution of sidelobe and spillover coupling to the measured gas emission.

During gas observations, the receiver views a cold background target located behind a base shroud, allowing molecular transitions to be observed in emission against a well-defined thermal background. All three targets contain a wedge-shaped absorbing cavity based on the calibration-target design developed for the European Space Agency's Arctic Weather Satellite \cite{Albers2024}. The absorber consists of a custom carbon- or iron-loaded epoxy compound which is mixed under vacuum with a high-speed planetary mixer to ensure homogeneous composition without gas-filled voids. The solid absorber is then machined to submillimeter tolerances to achieve low reflectivity near 557~GHz. The target design has been demonstrated to provide stable performance over a broad temperature range of 77--300~K. Each target is equipped with three PT1000 temperature sensors to provide accurate temperature monitoring and ensure high-precision radiometric calibration. The beam enters the absorbing cavity after reflection from a 90$^\circ$ folding mirror. The mirror assembly and adjacent target side walls are coated with a vacuum-compatible black coating developed by Acktar Ltd. to suppress stray infrared radiation.

A second flip mirror provides additional flexibility by switching the receiver view between two background targets at different temperatures or by directing the optical beam toward a transmitter for transmission spectroscopy measurements. Both flip mirrors are machined from aluminum and gold-coated.

\subsubsection{\label{sec:gas_cell} Gas Cell}
Heterodyne DSB radiometers down-convert signals from both the upper sideband (USB, RF frequencies above the effective LO frequency) and the lower sideband (LSB, RF frequencies below the effective LO frequency) into the same IF band. Consequently, the measured intensity of a spectral line originating in one sideband depends on the corresponding sideband gain ($G_{\mathrm{LSB}}$, $G_{\mathrm{USB}}$), which must be accounted for in the radiometer's absolute calibration. If the sideband gains are not accurately characterized, their uncertainty propagates directly into the retrieved line intensity. For example, one limitation of the MIRO instrument was the absence of a sideband-ratio correction, which introduced an uncertainty of approximately 20$\%$ in the measured line intensity \cite{Frerking2020}. For the WEEVIL radiometer, we constructed a dedicated gas cell to characterize the sideband gain ratio, following the calibration methods developed for the Herschel/HIFI instrument \cite{Higgins2011}.

\begin{figure}
 \centering
 \includegraphics[scale=0.06]{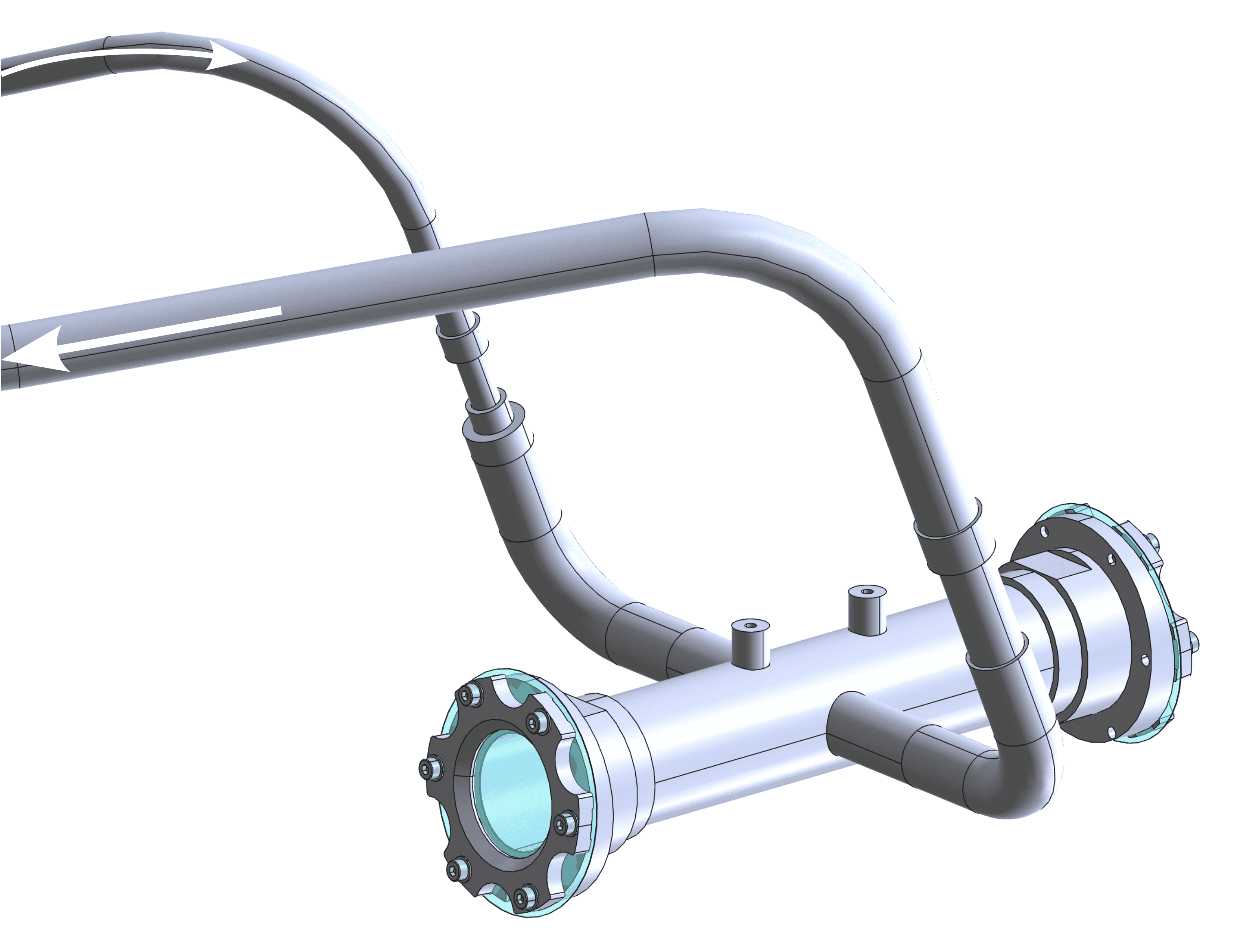}
 \caption{\label{fig:gas_cell_cad}
  Technical drawing of a gas cell.
  White arrows show the gas flow direction.}
\end{figure}

Fig.~\ref{fig:gas_cell_cad} shows the gas-cell design. The gas-cell flanges are tilted by 5$^\circ$ relative to the optical axis to reduce coupling of Fabry--Pérot standing waves back into the feedhorn. The cell has 30~mm apertures sealed with 25~$\mu$m Mylar windows, ensuring high transmission in the sub-mm wavelength range and allowing it to withstand sudden pressure differences. The gas cell is a compact stainless steel cylinder connected to the vacuum chamber flange via flexible stainless steel hoses. The internal surfaces of the gas cell cylinder are electropolished to reduce water sticking, which could otherwise cause pressure lags and additional outgassing at low pressures. 

A dedicated gas-injection system was developed to provide stable water vapor pressure inside the cell. To prevent excessive differential pressure across gas cell windows during initial evacuation of the main chamber, a bypass valve connecting the gas cell to the main chamber is kept open. Once the pre-vacuum is reached, the valve is closed, and the gas cell can be operated independently of the main chamber using a Pfeiffer HiCube 80 Eco turbo pumping station. The system allows freeze--pump--thaw cycles to remove dissolved gases from water, providing a purified water vapor source for spectroscopic calibration. The Pfeiffer CMR 364 capacitive pressure gauge measures the pressure inside the gas cell.      

\subsection{\label{sec:gas_model} Gas dynamics model}
\begin{figure*}
    \centering
    \makebox[0.5\textwidth][l]{(a)}\par
    \includegraphics[width=0.7\textwidth] {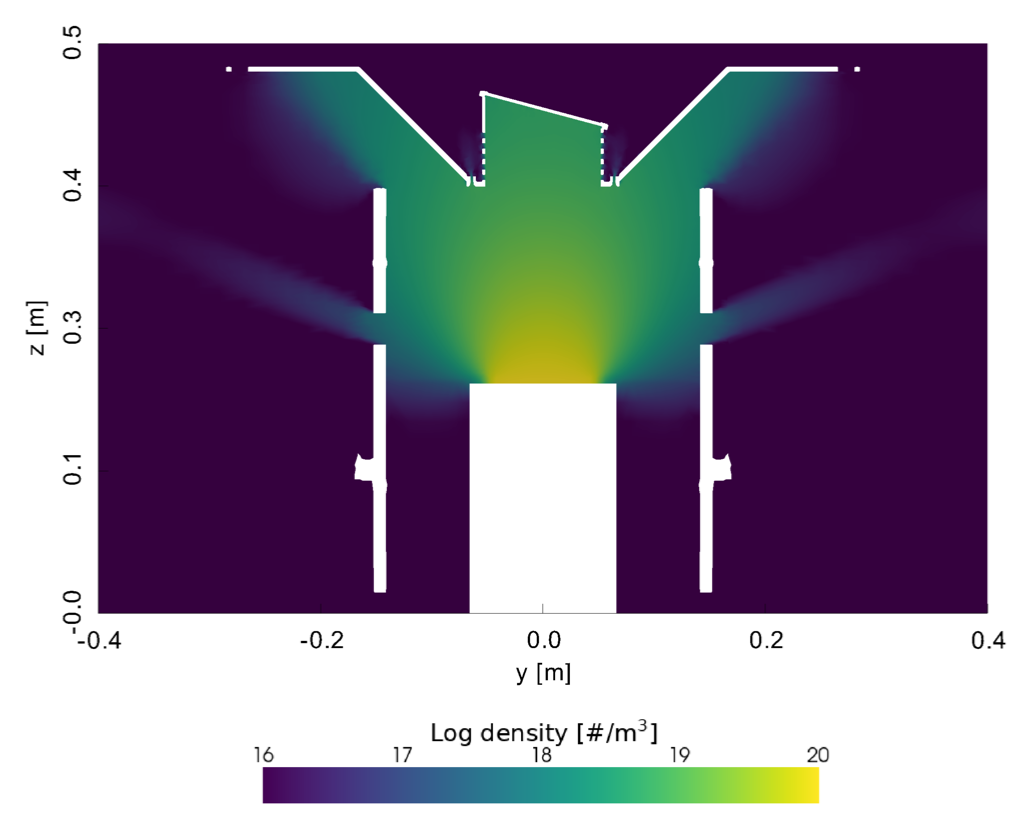}\par
    \medskip
    \makebox[0.55\textwidth][l]{(b)}\par
    \includegraphics[width=0.7\textwidth]{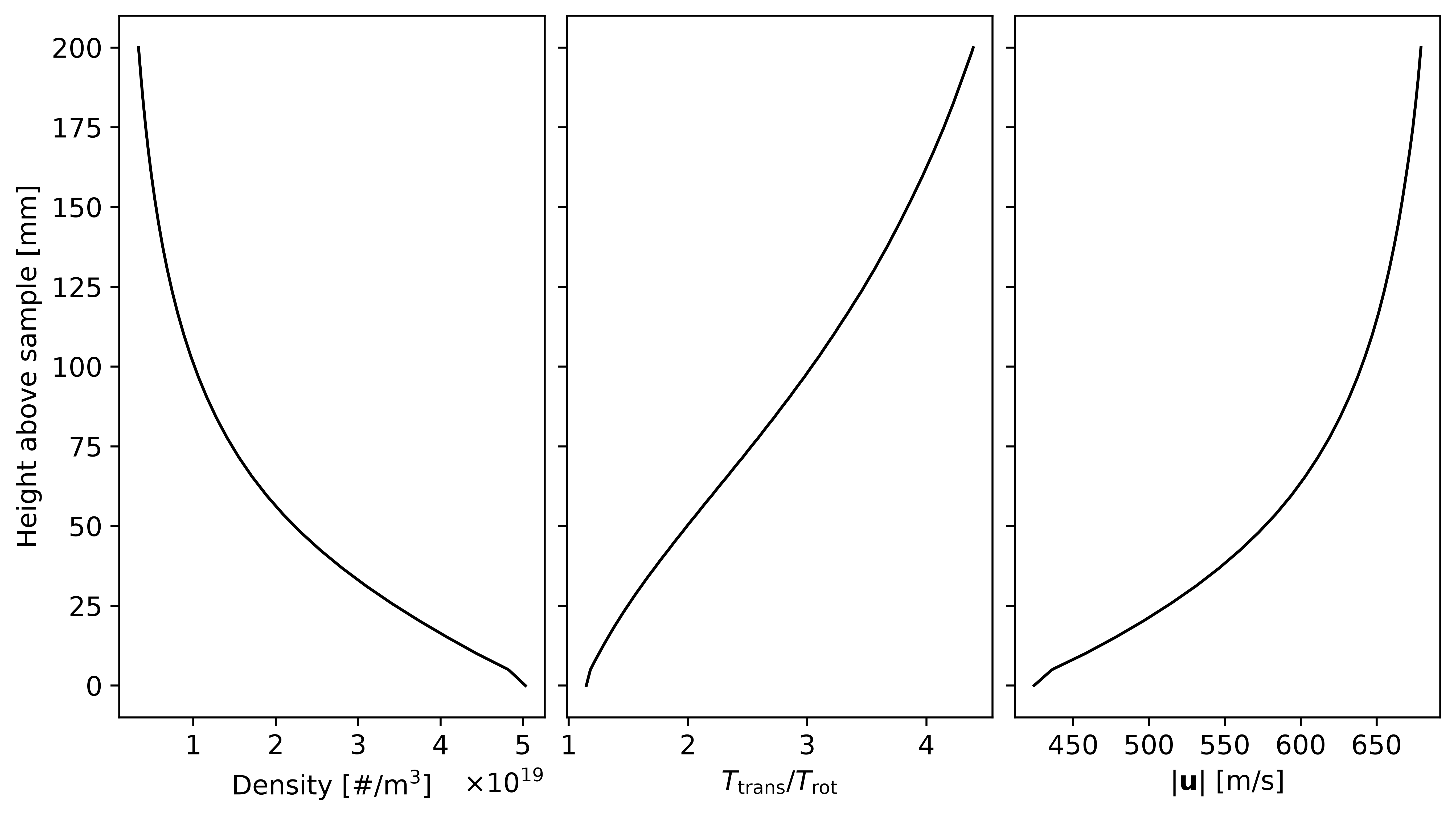}\par
    \caption{\label{fig:dsmc_model}
    Results of the 3D DSMC simulation of gas expansion inside the vacuum
    chamber during the sublimation experiment.
    Top: Simulated H$_2$O number-density field in a vertical slice through
    the chamber.
    Bottom: Vertical profiles of the H$_2$O number density,
    translational-to-rotational temperature ratio, and gas velocity
    magnitude along the central plume axis.
    The simulation contains approximately 200,000 computational cells.
    The base and top shrouds (Fig.~\ref{fig:chamber_drawing}) are modeled as diffuse walls at 100~K, with a
    reflection probability of 0.05. The sublimating surface is maintained
    at 200~K, and the molecular number density is prescribed according to
    the Hertz--Knudsen relation as
    $5.89\times10^{19}$~m$^{-3}$.}
\end{figure*}

We developed a Direct Simulation Monte Carlo (DSMC) model of the experimental setup to support vacuum-chamber design and interpret the radiometric measurements. The model is based on the open-source dsmcFoamPlus solver \cite{White2018}, implemented within the OpenFOAM framework. It enables simulations of rarefied gas flow inside the vacuum chamber using appropriate gas--surface boundary conditions and accounts for energy exchange between the translational and rotational degrees of freedom of gas molecules. Fig.~\ref {fig:dsmc_model} shows an example DSMC simulation of the gas plume inside the experimental chamber. Simulations show that gas released from the sublimating sample is under non-local thermodynamic equilibrium (NLTE) conditions, with the rotational temperature ($T_\mathrm{rot}$) deviating significantly from the translational temperature ($T_\mathrm{trans}$) as the gas expands into the vacuum chamber. Therefore, the radiative transfer model used in data processing should account for NLTE conditions.

\section{Commissioning}
After completing the system assembly, we conducted a series of commissioning tests to characterize performance and assess compliance with the requirements defined in Sec.~\ref{sec:setup_requirements}.

\subsection{\label{sec:pressure_temperature_commissioning} Pressure and temperature}
To characterize the system's background gas load, we isolated the vacuum chamber from the pumping system. The resulting pressure-rise rate, which includes contributions from outgassing, desorption, and possible residual leakage, corresponds to a total gas load of approximately $1\times10^{-3}$~mbar~L/s. Using the TMP, the chamber pressure reaches $\approx 1.5\times10^{-6}$~mbar after 40~h of continuous pumping. The cooled shrouds reduce the pressure to $\approx 8-9\times 10^{-7}$~mbar, while the sample cryocooler allows the chamber to reach below $1\times 10^{-7}$~mbar.

Table~\ref {tab:chamber_temperature} presents the physical temperatures of the components inside the vacuum chamber during spectroscopic measurements. The radiometric components maintain thermal stability within $\pm$ 0.3~K/h. This low drift allows the long integration times required for radiometer measurements without sudden temperature changes. The sample attached to the cryocooler can be cooled to temperatures as low as 40~K. PID-regulated heating elements allow the sample temperature to increase gradually to 220~K at a predefined heating rate. After reaching steady state, the sample temperature typically remains within $\pm0.5$~K of the set point, with short-term oscillations of approximately 0.1--0.2~K.

\begin{table}[ht]
\centering
\caption{Steady-state temperatures and temperature stability of the main
components. $T$ is the mean temperature, while $\sigma_{10\,\mathrm{min}}$
and $\sigma_{1\,\mathrm{h}}$ are the temperature standard deviations over
10~min and 1~h periods, respectively.}
\vspace{3mm}
\label{tab:chamber_temperature}
\begin{tabular}{lccc}
\hline
Component & $T$ [K] & $\sigma_{10\,\mathrm{min}}$ [K] &
$\sigma_{1\,\mathrm{h}}$ [K] \\
\hline
Hot target        & 274.77 & 0.23 & 1.18 \\
Cold target       & 87.17  & 0.08 & 0.19 \\
Background target & 85.29  & 0.11 & 0.23 \\
Mixer             & 102.78 & 0.10 & 0.34 \\
LNA               & 98.43 & 0.09 & 0.24 \\
Base shroud       & 84.79  & 0.53 & 0.67 \\
Top shroud        & 88.69  & 0.09 & 0.37 \\
QCM               & 86.02  & 0.12 & 0.50 \\
\hline
\end{tabular}
\end{table}

\subsection{\label{sec:scanning_mechanism_commissionning} Scanning mechanism}

The scanning mechanism was extensively tested over its full 200~mm vertical travel range under both ambient and high-vacuum conditions, including operation with the cryocooler cold head at its operating temperature. During the prototyping phase, the linear-stage mechanism underwent load testing, which involved repeatedly moving a 150~kg deadweight vertically to verify sufficient torque margin in the stepper motor and ball-screw drive. The system's "zero position" is defined as a 5~mm vertical distance between the radiometer beam axis and the ice-sample surface, corresponding to the $-10$~dB beam radius (Fig.~\ref{fig:beam_profile_grasp}). A Hall sensor determines the zero position.  

We then tested the scanning mechanism during sublimation experiments. Starting from the zero position, the sample was moved progressively downward, increasing the distance between the ice surface and the fixed radiometer beam and enabling measurements of the vertical gas-plume profile. Sec.~\ref{sec:results} presents the obtained spectra.

\subsection{\label{sec:radiometer_commissioning} Radiometer calibration}

For radiometer heterodyne receiver calibration, we conducted a dedicated measurement campaign in a vacuum chamber with cryogenic cooling, while we characterized the radiometer optics in a quasi-optical laboratory at the Institute of Applied Physics (IAP), University of Bern. We performed gas cell measurements in the WEEVIL chamber.  

\subsubsection{\label{sec:sensitivity} Sensitivity} 

Eq.~\ref{eq:snr_radiometer} gives the ideal radiometer-limited Signal-to-Noise Ratio (SNR). In frequency-switching observations (discussed below), differencing increases the noise by a factor of $\sqrt{2}$, reducing the SNR accordingly \cite{Gulkis2007}. Calibration noise further reduces the SNR.

\begin{equation}
\label{eq:snr_radiometer}
\mathrm{SNR}
=
\frac{G_s\,T_{\mathrm{line,SSB}}}
{T_{\mathrm{sys,DSB}}}
\sqrt{B\tau},
\end{equation}

where $B$ is the bandwidth of the spectrometer channel (24~kHz), $\tau$ is the integration time, $T_{\mathrm{line, SSB}}$ is the brightness temperature of the spectral line in a single sideband, $T_{\mathrm{sys, DSB}}$ is the noise temperature of the DSB system, including the noise temperature of the DSB receiver and the antenna temperature ($T_{\mathrm{a}}$), and $G_s$ is the normalized gain of the sideband containing the spectral line. The IF spectrum of a DSB receiver contains contributions from both the lower sideband (LSB) and the upper sideband (USB), and a spectral line may originate from either sideband.

For sideband ratio measurements, we introduced a second background target maintained at room temperature (hot target), following the sideband ratio calibration procedure used elsewhere \citep{Higgins2011, Neumaier2014}. First, we performed reference measurements with the evacuated gas cell while alternately observing the hot and cold background targets. We then repeated the measurements with water vapor introduced into the gas cell at pressures at which the 556.9~GHz transition becomes optically thick (line saturation). As a result, sideband gains can be calculated directly using $G_{\mathrm{LSB}}+G_{\mathrm{USB}}=1$. The measured sideband gain is $G_\mathrm{LSB} = 0.502 \pm 0.006$, which indicates equal gain in the two sidebands. 

We determined the receiver noise temperature using the Y-factor method. In this technique, the radiometer alternately observes calibration cold and hot targets. The measured output powers, along with the targets' known physical temperatures, are used to calculate the receiver noise temperature. During the calibration campaign, the LO power was adjusted to minimize the receiver noise temperature, and the optimal settings were stored in a look-up table. 

\begin{figure}
\includegraphics[scale=0.5]{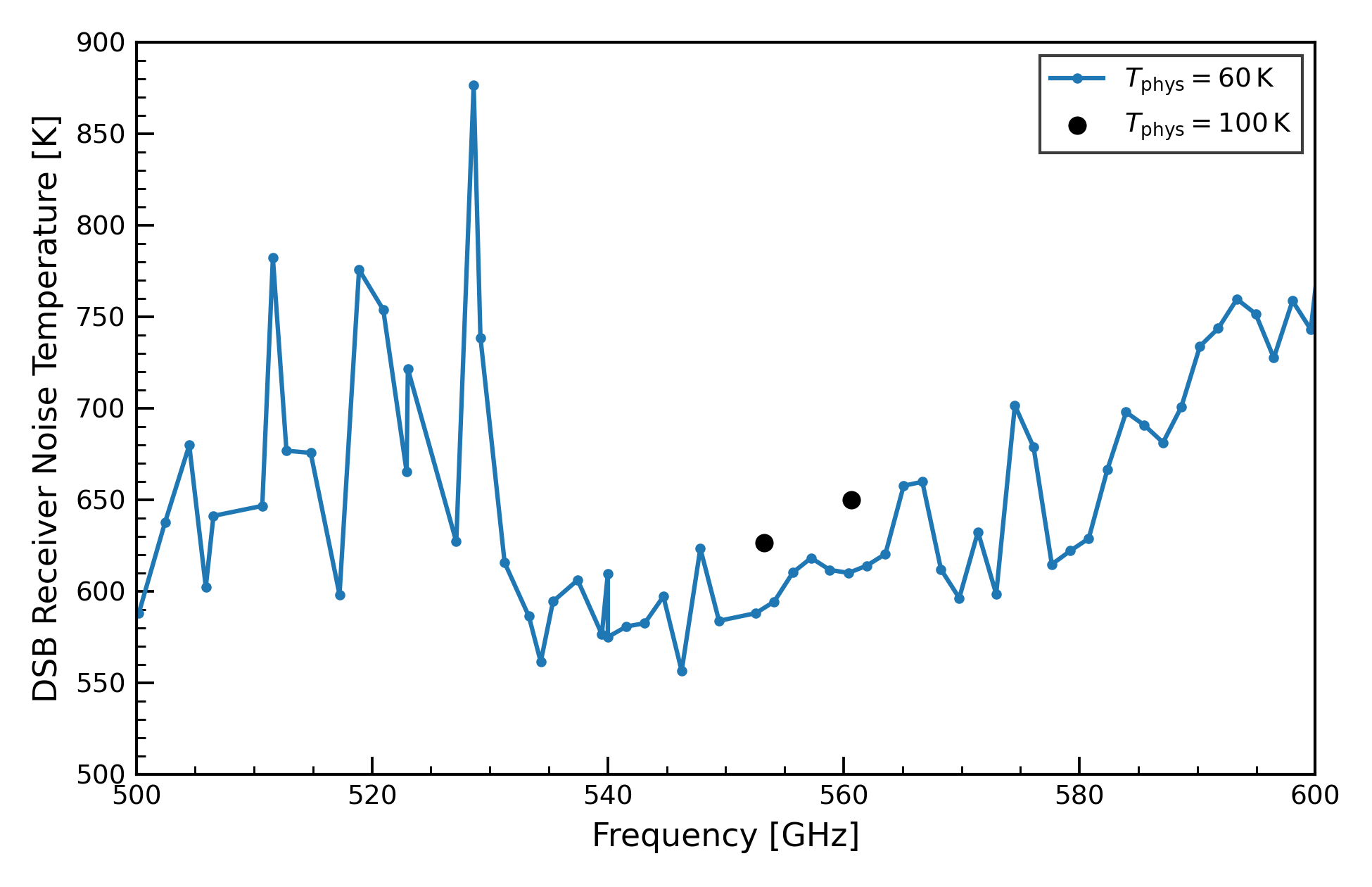}
\caption{\label{fig:receiver_noise_temperature} Double-sideband (DSB) noise temperature of the radiometer measured at two mixer physical temperatures. The blue curve corresponds to a mixer physical temperature of 60~K, and the black dots to 100~K.}
\end{figure}

Fig.~\ref{fig:receiver_noise_temperature} presents the measured DSB receiver noise temperatures. During the calibration campaign, the radiometer achieved a DSB receiver noise temperature of 612~K at 560~GHz with the mixer operating at a physical temperature of 60~K. In the WEEVIL vacuum chamber, the mixer operates at about 100~K and achieves a DSB receiver noise temperature of 650~K at 560~GHz. 

Using the obtained DSB receiver noise temperatures and Eq.~\ref{eq:snr_radiometer}, we can calculate the integration times required to achieve K and sub-K resolution. As shown in Table~\ref{tab:radiometer_integration_time}, WEEVIL has a lower noise temperature than the other flight radiometers considered here, except HIFI, which used a cryogenically cooled SIS mixer for sub-mm detections \cite{deGraauwTh2010}.

In addition to thermal noise (Eq.~\ref{eq:snr_radiometer}), time-dependent gain fluctuations ($\Delta G$) affect radiometer sensitivity and arise from changes in physical temperature, bias voltages, or amplifier gain. At sufficiently long integration times, these fluctuations can dominate thermal noise, limiting the radiometer's effective integration time. The WEEVIL radiometer's maximum integration time is 100~seconds for a 24.4~kHz spectrometer channel. For observations of narrow spectral lines, however, frequency switching can suppress slowly varying gain fluctuations and extend the effective integration time.

We calibrate the spectra obtained with the WEEVIL radiometer using an approach similar to that used for the MIRO sub-mm receiver \citep{Frerking2020}. Briefly, spectra recorded at two different LO frequencies are subtracted to form a differential spectrum, which is further calibrated using measurements of hot--cold calibration targets. The calibration accounts for the fact that the emission line originates from one sideband (e.g., LSB), whereas the continuum emission from the calibration targets is detected in both sidebands. Finally, the calibrated difference spectrum is folded by aligning and averaging the positive and negative line features produced by the frequency switching.

We tested the frequency-switching technique with a 30~s switching interval, with hot/cold calibration measurements performed every 10~min. We observed no significant changes in the spectral baseline between successive calibrations using a 5~MHz frequency-switching offset. 

\begin{table}[ht]
\centering
\caption{Integration time required to detect a 556.9~GHz line of $\Delta T = 1, 0.1$ K with $\mathrm{SNR}=3$, $B=44$ kHz, and $T_{\mathrm{a}}=100$ K using Eq.~\ref{eq:snr_radiometer}.}
\vspace{3mm}
\label{tab:radiometer_integration_time}
\begin{tabular}{lrrr}
\hline
Instrument & $T_{\mathrm{rec}, \mathrm{DSB}}$ [K] & 
$\tau_{\mathrm{1K, SNR=3}}\,\mathrm{[min]}$ &
$\tau_{\mathrm{0.1K, SNR=3}}\,\mathrm{[h]}$ \\
\hline
WEEVIL & 650 & 7.7 & 12.8 \\
SWI (Juice) & 1500 \footnote{\citet{Hartogh2026}} & 34.9 & 58.2 \\
MIRO (Rosetta) & 3597 \footnote{\citet{Gulkis2007}} & 186.4 & 310.6 \\
SMR (Odin) & 1650 \footnote{\citet{Nordh2003}} & 41.8 & 69.6 \\
HIFI (Herschel) & 100 \footnote{\citet{deGraauwTh2010}} & 0.5 & 0.9\\
\hline
\end{tabular}
\end{table}

\subsubsection{\label{sec:beam_profile}Beam profile and targets reflectance}

\begin{figure}[t]
    \centering
    \begin{minipage}{0.9\linewidth}
        \raggedright
        (a)\par
        \includegraphics[width=\linewidth]{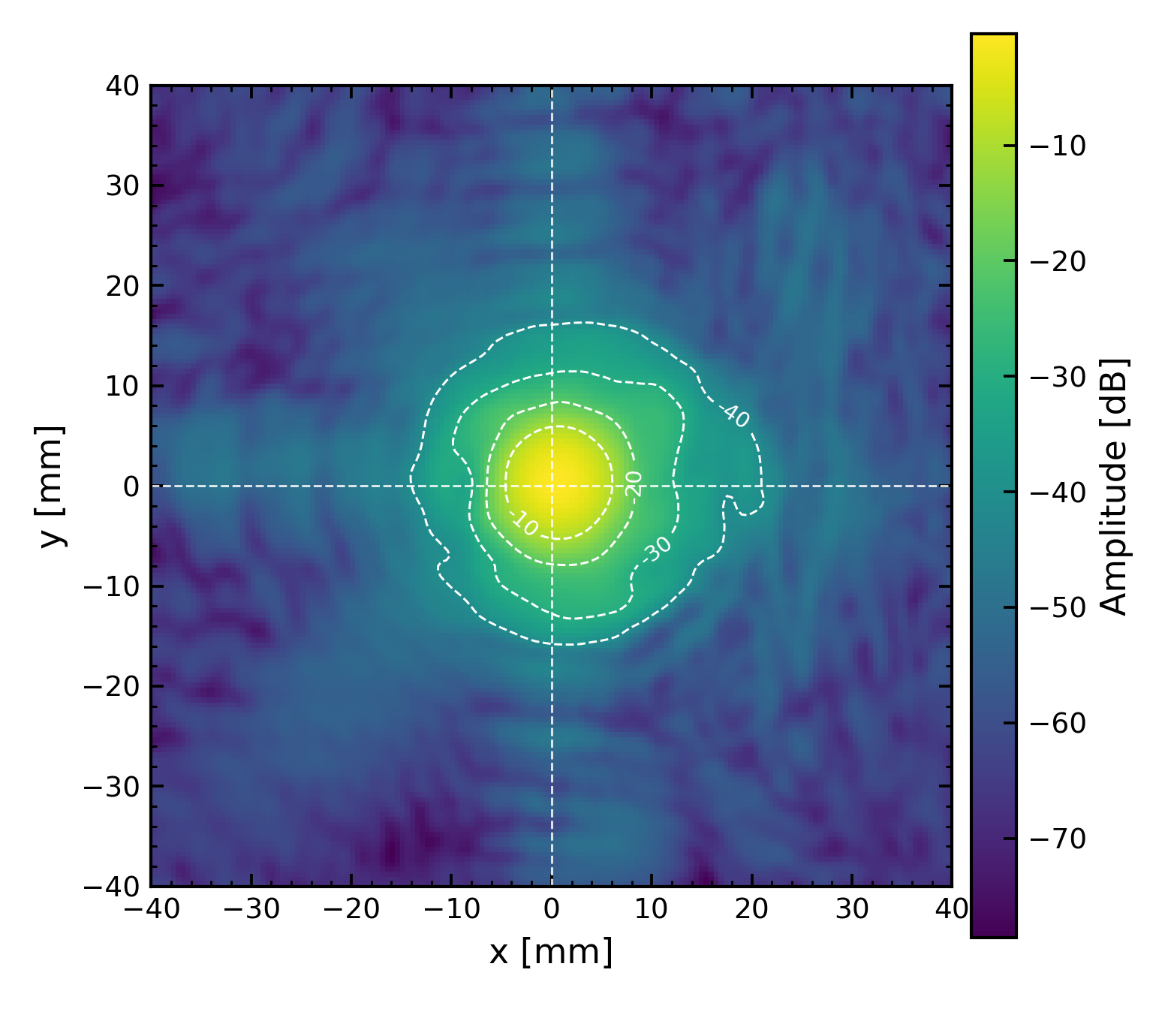}
    \end{minipage}\par\medskip
    \begin{minipage}{0.9\linewidth}
        \raggedright
        (b)\par
        \includegraphics[width=\linewidth]{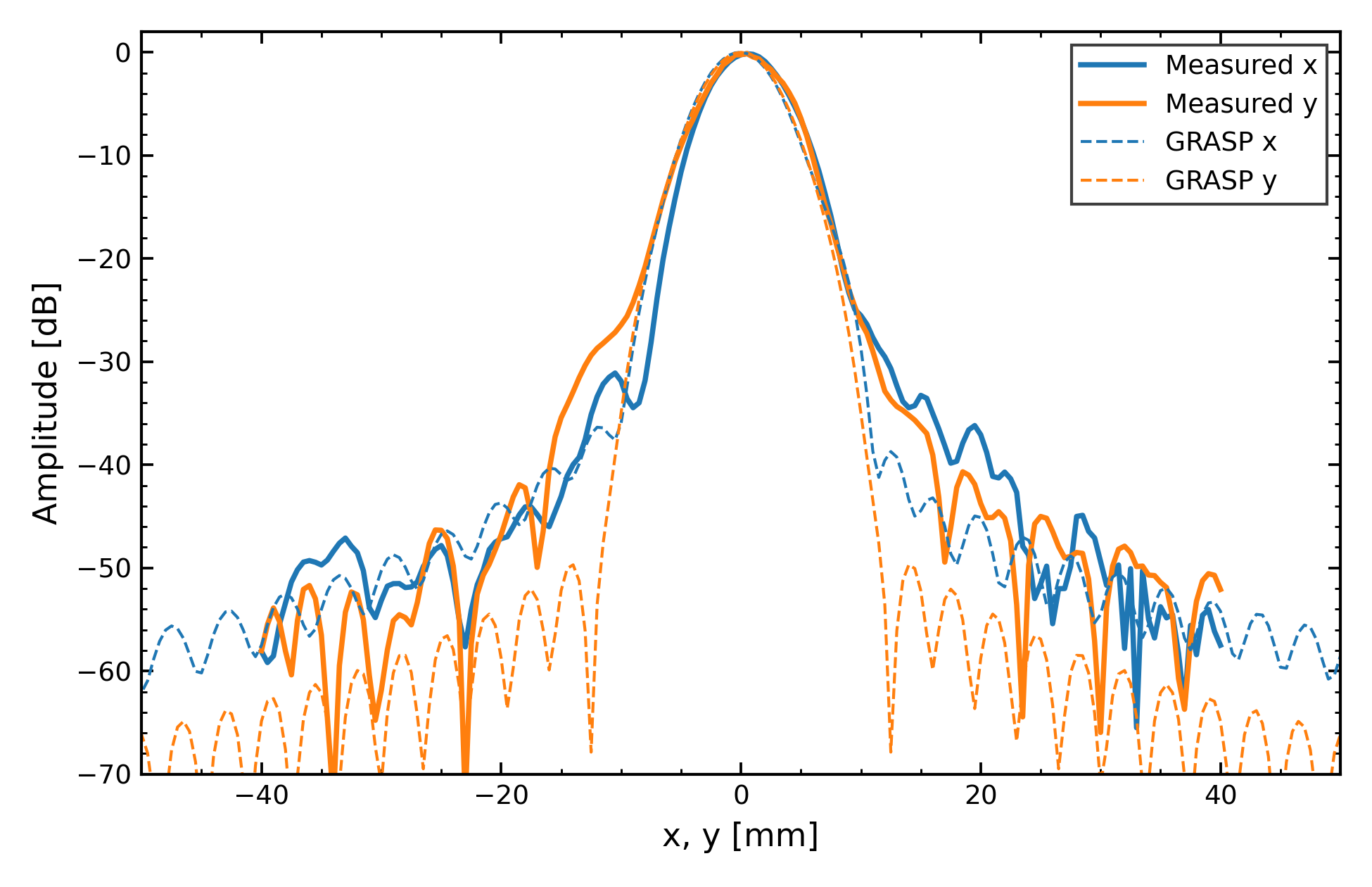}
    \end{minipage}

    \caption{\label{fig:beam_profile_grasp}
    Beam-pattern measurements of the 557~GHz radiometer at the expected
    center of the gas plume. (a) Normalized two-dimensional beam-amplitude
    profile. (b) Beam-profile cuts along the $x$ and $y$ axes compared
    with GRASP simulations.}
\end{figure}

To characterize the antenna beam pattern, we conducted dedicated near-field optical measurements. The submillimeter source of the ABmm vector network analyzer (VNA) was mounted on a planar $xy$ scanning stage. We positioned the WEEVIL radiometer mirror block (Fig.~\ref{fig:radiometer_layout}) 268.4~mm from the probe horn, corresponding to the focus of the radiometer antenna Gaussian beam at the center of the base shroud. The measured near-field pattern has a radius of approximately 10~mm at -10~dB relative to the beam center, with sidelobe levels below -30~dB. As shown in Fig.~\ref{fig:beam_profile_grasp}, the measured beam pattern agrees well with the optical simulations, with a higher measured response than the simulations in the $y$-cut outside the main beam, which can be attributed to misalignment and scattering in the test setup.  

We measured the coherent backscattering coefficient ($S_{11}$) of the calibration targets in the range 500--600~GHz using the method described by \citet{Jacob2018}. The measured $S_{11}$ is below $-50$~dB, indicating very low reflectance and thus minimizing spectral baseline ripples from reflections off the calibration targets.

\subsection{\label{sec:qcm_commissioning} QCM}

\begin{figure}
 \includegraphics[scale=0.5]{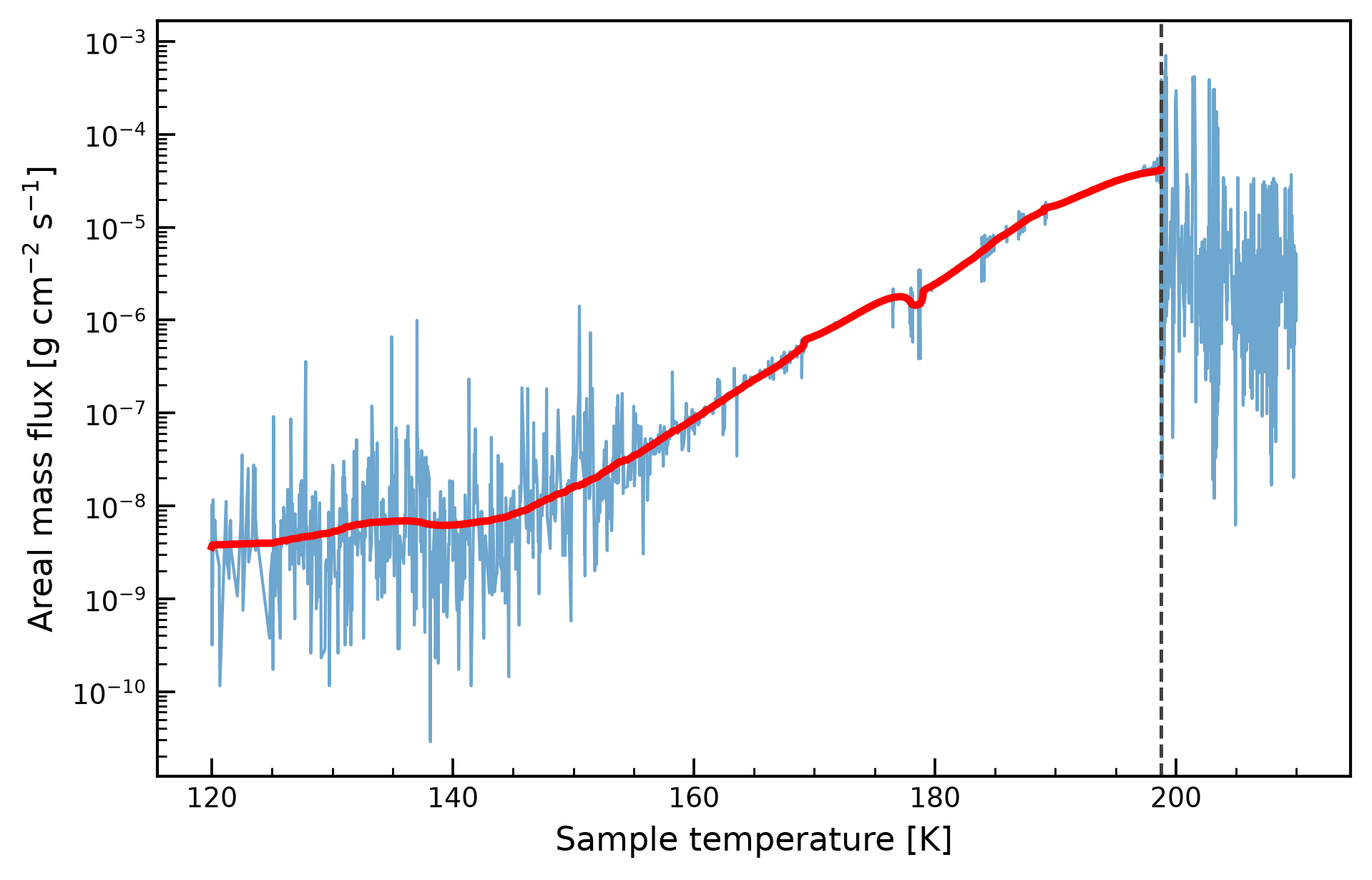}
 \caption{\label{fig:areal_mass_flux_qcm} Areal mass fluxes measured at different temperatures of the water-ice sample. The blue curve shows the mass flux calculated using Eq.~\ref{eq:qcm_equation}, while the red curve shows the smoothed values. The grey dashed line indicates the onset of the QCM instability.}
\end{figure}

For initial sublimation experiments, we attached a quartz crystal microbalance sensor to an aluminum plate mounted on the bottom of the top shroud (Fig.~\ref{fig:chamber_drawing}, \ref{fig:chamber_in_lab}) to measure the mass flux from the sublimating water-ice sample. We monitored the Mark-10 QCM temperature using a PT100 sensor integrated into the QCM head. During operation, the QCM temperature was approximately 88~K with stability about 0.1~K (Table~\ref{tab:chamber_temperature}). According to the Sauerbrey equation, the change in the quartz crystal's resonance frequency, $\Delta{f}$, is directly proportional to the deposited mass, $\Delta{m}$. For the Mark-10 QCM, this relation is as follows:    

\begin{equation}
 \label{eq:qcm_equation}
 \Delta{f}
 =
 -2.26\times 10^{-6}f^{2}\Delta{m}/A,
\end{equation}

where $f$ is the quartz crystal resonance frequency (15~MHz) and $A$ is the crystal area exposed to the incident mass flux (0.316 $\mathrm{cm}^{2}$). Because the QCM operates at a low temperature (88~K), re-sublimation of deposited water molecules is negligible, and we assume a unity sticking coefficient, so the measured mass-deposition rate equals the incident mass flux. Fig.~\ref{fig:areal_mass_flux_qcm} shows the measured areal mass flux for different ice-sample temperatures. The QCM frequency readings become unstable when the deposited ice layer exceeds approximately 6~$\mu$m, corresponding to a frequency shift of approximately 0.29~MHz. According to Fig.~\ref{fig:areal_mass_flux_qcm}, the QCM sensor can reliably measure the mass flux from the sublimating sample for ice temperatures up to 200~K.    

\section{\label{sec:results}Results}

To assess the functionality of the complete experimental setup, we performed several tests with a sublimating water-ice sample. During these tests, the temperature of the sample holder was used as a proxy for the ice temperature. The sample temperature was controlled using resistive heaters integrated into the sample holder. The implemented heating method does not fully reproduce conditions on comet surfaces, where solar irradiation is the primary heat source. Furthermore, as the ice-surface temperature approaches 180--200~K, the sample-holder temperature deviates significantly from the surface temperature. Because the primary objective of this work is to demonstrate and characterize the capabilities of the experimental setup, we do not investigate these effects in detail here; we will address them in future experiments by directly measuring the ice bulk and surface temperatures. 

\begin{figure}[t]
    \centering
    \begin{minipage}{0.9\linewidth}
        \raggedright
        (a)\par
        \includegraphics[width=\linewidth]{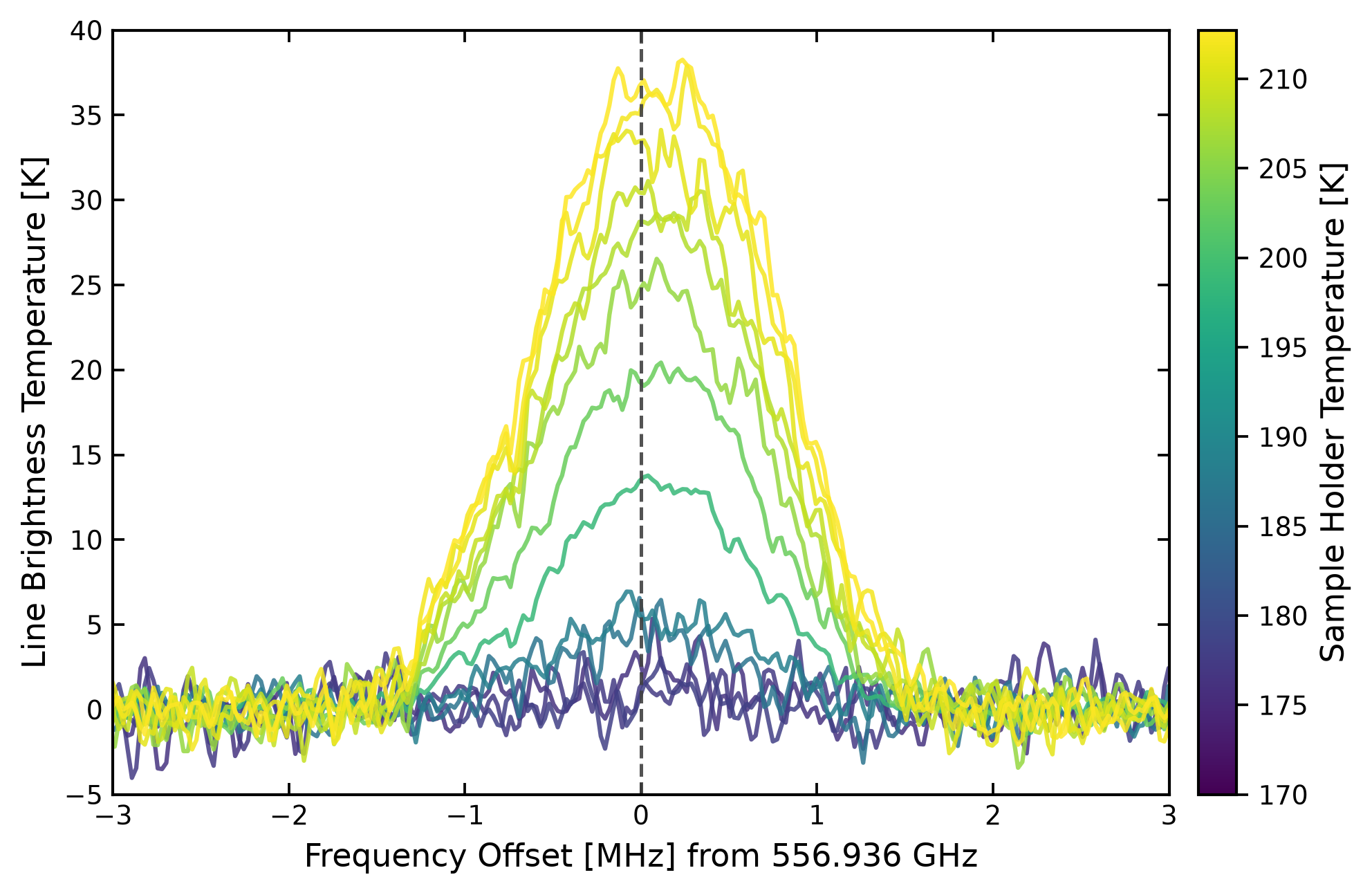}
    \end{minipage}\par\medskip
    \begin{minipage}{0.9\linewidth}
        \raggedright
        (b)\par
        \includegraphics[width=\linewidth]{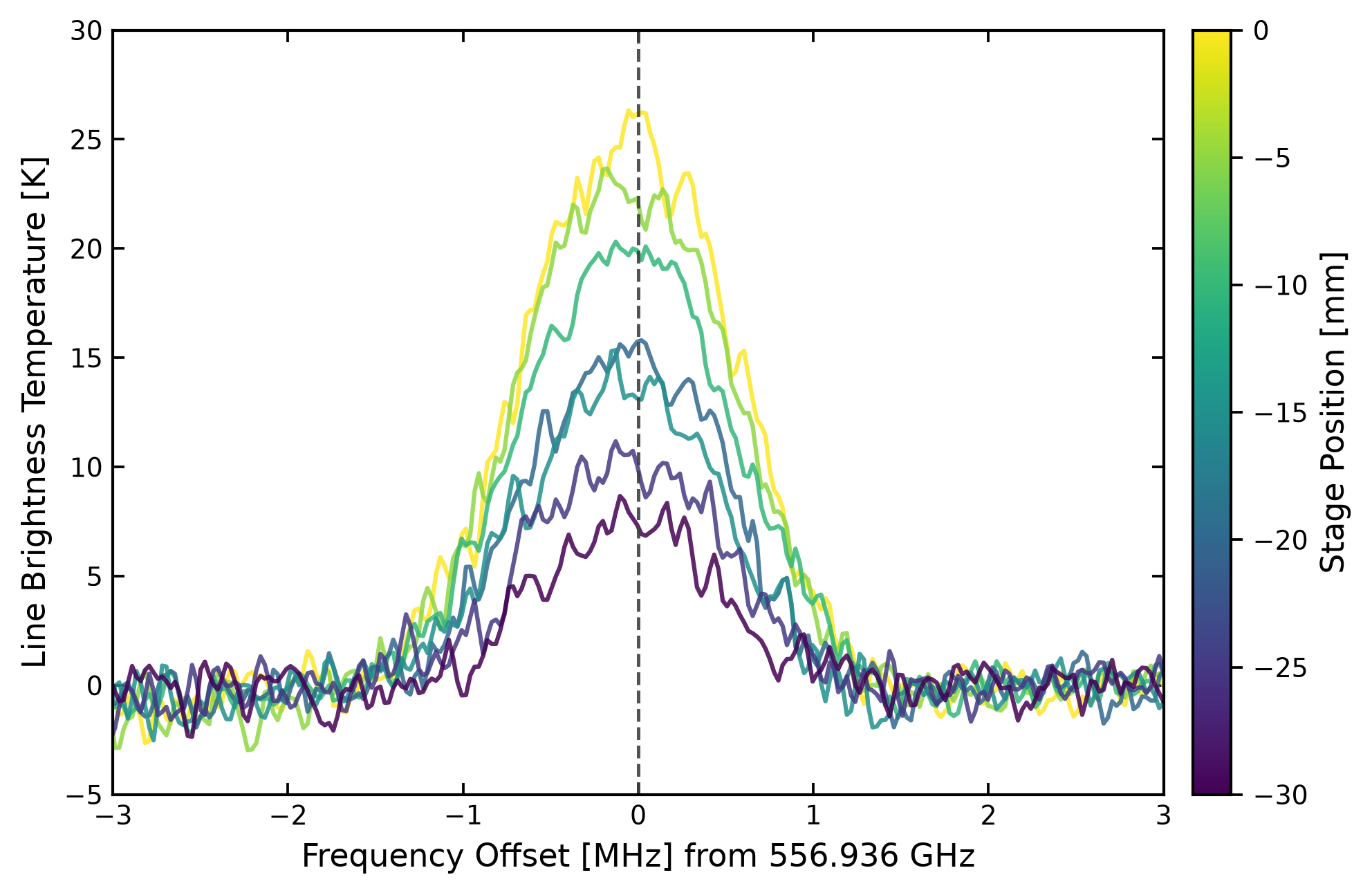}
    \end{minipage}

    \caption{\label{fig:spectra_1}
    Water-vapor ground-state transition measured with the WEEVIL
    radiometer during water-ice sublimation. The measurements used
    frequency switching with a 5~MHz separation between the two LO
    settings. (a) Spectra at different sample-holder temperatures,
    with integration times of 2--20~min. (b) Spectra at different
    vertical positions of the scanning stage relative to its zero
    position, at a sample-holder temperature of 208~K and an
    integration time of 10~min.}
\end{figure}

In the first test, we gradually heated the sample to 210~K while keeping the radiometer beam position relative to the sample surface constant. As shown in Fig.~\ref{fig:spectra_1}, the sub-mm signal from the water vapor line begins to dominate over the background at sample holder temperatures above $\approx$185~K. 

The second test evaluated the scanning mechanism and the setup's ability to scan the gas plume. During measurements, we progressively moved the sample holder downward, away from the radiometer beam axis. We observed the expected decrease in emission-line intensity with increasing distance from the ice surface (Fig.~\ref{fig:spectra_1}).  

\begin{figure}
 \includegraphics[scale=0.5]{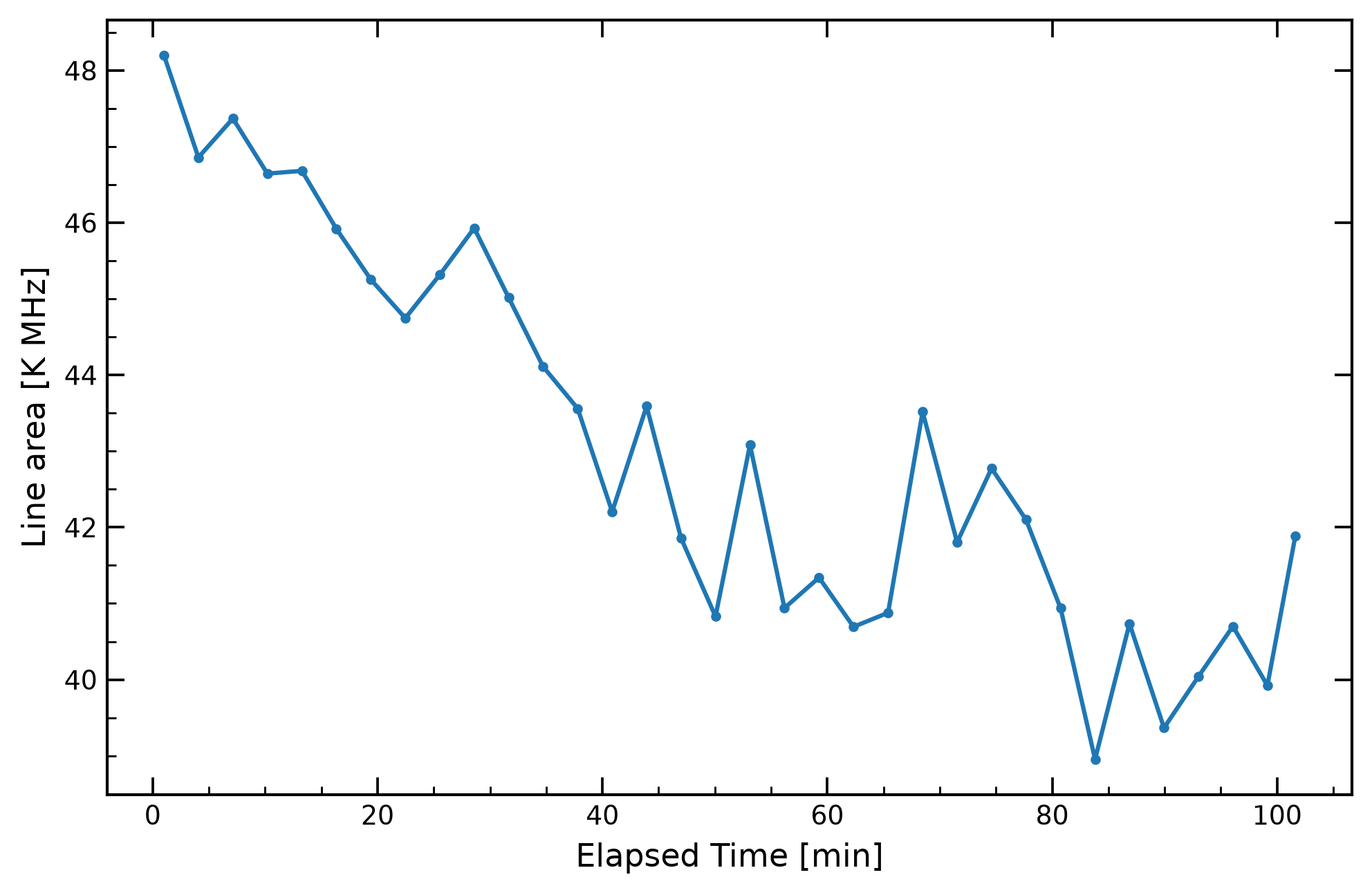}
 \caption{\label{fig:line_temporal_changes} Temporal evolution of the water vapor line (556.9~GHz) area (brightness temperature integrated over frequency) during sublimation of the water-ice sample at a constant sample-holder temperature of 208~K.}
\end{figure}

In the third test, we maintained the sample holder temperature at 208~K for 100~min. The water vapor emission-line area (strength) decreased gradually over time (Fig.~\ref{fig:line_temporal_changes}), showing the setup's sensitivity to temporal changes in the gas above the sublimating sample. The origin of this decrease remains undetermined and may involve changes in the ice-surface temperature, gas density, or collisional conditions above the sample.

\section{\label{sec:conclusion} Conclusion}

We introduced a new experimental setup, WEEVIL, designed to investigate in detail gas released from sublimating ice samples under comet-- and icy-moon-relevant conditions. For the first time, a sub-mm radiometer is used in a laboratory facility to investigate rarefied gas flow above a sublimating ice surface. In parallel, we developed a gas-dynamics model to characterize the evolving gas flow and support interpretation of the radiometric measurements. We demonstrated the complete setup's functionality during initial water-ice sublimation experiments, during which we successfully acquired the first sub-mm spectra.

We summarize the main capabilities of the setup as follows: 

\begin{itemize}
 
 \item \textit{Cooling}. The LN$_2$-cooled shrouds cool the radiometer and auxiliary sensors, including the QCM, to approximately 85~K. The cryocooler controls the ice-sample temperature from 40 to 300~K.
 
 \item \textit{Vacuum}. The vacuum system reaches an ultimate chamber pressure below $10^{-7}$~mbar.
 
 \item \textit{Scanning mechanism}. The sample holder can be vertically translated in a range of 200~mm, allowing measurements of the gas plume at different distances from the ice surface.
 
 \item \textit{Radiometer}. The heterodyne DSB radiometer operates around the ground-state transition of water vapor at 556.9~GHz and achieves a DSB receiver noise temperature of approximately 650~K under WEEVIL operating conditions. The receiver has an Allan stability time of approximately 100~s. The antenna beam has a diameter of approximately 10~mm at the $-10$~dB level in the center of a gas plume. The calibration targets have reflectance below $-50$~dB around the operating frequency. The measured sideband ratio is 0.502:0.498. The radiometer supports both passive and active measurements. In the latter case, it uses a sub-mm transmitter.
 
 \item \textit{Gas sensors}. The setup includes pressure gauges and a quartz crystal microbalance (QCM) to provide complementary data for radiometric observations. The QCM enables measurements of the mass flux originating from the sublimation of water ice at sample temperatures of up to 200~K.  
 
 \item \textit{Gas cell and gas-injection system}. A dedicated gas cell can be installed inside the vacuum chamber for spectroscopic calibration and radiometer characterization. The gas-injection system provides a controlled supply of gases and gas mixtures to either the gas cell or the cooled sample holder for deposition.
 
\end{itemize}
 
The commissioning demonstrates that WEEVIL can resolve the spectral, spatial, and temporal evolution of gas profiles originating from sublimating ice under controlled laboratory conditions. Future experiments will combine radiometric measurements with DSMC gas-dynamics simulations and radiative-transfer modeling to investigate the sublimation process and constrain the subsurface microphysical properties of sublimating ices. Particular emphasis will be placed on experiments with porous dust layers to characterize their influence on gas profiles. These measurements will provide laboratory constraints for theoretical models of cometary outgassing and support the interpretation of MIRO observations. The system is also relevant to water vapor emission in plumes from sub-surface oceans on icy moons. 

% References should be done using the \cite, \ref, and \label commands
%\section{}
%\label{}
%\subsection{}
%\subsubsection{}

% If in two-column mode, this environment will change to single-column format so that long equations can be displayed. 
% Use only when necessary.
%\begin{widetext}
%$$\mbox{put long equation here}$$
%\end{widetext}

% Figures should be put into the text as floats. 
% Use the graphics or graphicx packages (distributed with LaTeX2e).
% See the LaTeX Graphics Companion by Michel Goosens, Sebastian Rahtz, and Frank Mittelbach for examples. 
%
% Here is an example of the general form of a figure:
% Fill in the caption in the braces of the \caption{} command. 
% Put the label that you will use with \ref{} command in the braces of the \label{} command.
%
% \begin{figure}
% \includegraphics{}%
% \caption{\label{}}%
% \end{figure}

% Tables may be be put in the text as floats.
% Here is an example of the general form of a table:
% Fill in the caption in the braces of the \caption{} command. Put the label
% that you will use with \ref{} command in the braces of the \label{} command.
% Insert the column specifiers (l, r, c, d, etc.) in the empty braces of the
% \begin{tabular}{} command.
%
% \begin{table}
% \caption{\label{} }
% \begin{tabular}{}
% \end{tabular}
% \end{table}

% If you have acknowledgments, this puts in the proper section head.
\begin{acknowledgments}
This work was supported through a MARVIS (Multidisciplinary Advanced Research Ventures in Space) program of the Swiss Department for Business, Education, and Research (SBFI) called SUBICE. SUBICE is a project of the University of Bern (UniBe), the University of Applied Sciences and Arts, Western Switzerland (HES-SO), and Thales-Alenia Space Switzerland (TASCH). The authors thank the Space Research $\&$ Planetary Sciences fine mechanical and electronics departments, and specifically thank Alexander Schläfli, Sacha Häusler, Iljadin Manurung, Florian Hostettler and Harald Mischler for their help in constructing the setup. Furthermore, authors thank Tobias Plüss and Mikko Kotiranta (both IAP, University of Bern), and Jeffrey Hesler (VDI) for their technical support in the development of the radiometer. The authors also thank André Galli and Raphael Hänggi for their support with the quartz crystal microbalance sensor, Niels Ligterink for the gas-injection system for the gas cell, and Mathias Brändli, Antoine Pommerol and Rafael Ottersberg for their support with the vacuum chamber design.
\end{acknowledgments}

\section*{\label{sec:declarations} author declarations section}
\subsection*{Conflict of Interest}
The authors have no conflicts to disclose.

\section*{Data Availability}
The data that support the findings of this study are available from the corresponding author upon reasonable request.

% Create the reference section using BibTeX:
\bibliography{weevil_commissioning}

\end{document}